\documentclass[journal, onecolumn]{IEEEtran}

\usepackage{ifpdf}
\usepackage{xcolor}
\usepackage{textcomp}
\usepackage{leading}
\usepackage{calligra}
\definecolor{salmon}{RGB}{250,128,114}
\definecolor{lightsteelblue}{RGB}{176,196,222}
\usepackage{dutchcal}
\usepackage{calrsfs}
\usepackage{mathrsfs}
\usepackage[mathscr]{eucal}

\usepackage{cite}

\ifCLASSINFOpdf
   \usepackage[pdftex]{graphicx}
\else
   \usepackage[dvips]{graphicx}
\fi

\usepackage{amsmath}
\usepackage{amsthm}
\usepackage{amsfonts}
\usepackage{amssymb}
\usepackage{graphicx}
\usepackage{calrsfs}
\usepackage{mathrsfs}
\usepackage{multirow}
\usepackage[mathscr]{eucal}
\usepackage{mathtools}

\usepackage{algorithm}
\usepackage{algorithmic}

\usepackage{array}
\usepackage{mdwtab}
\usepackage{tabularray}
\usepackage{tabularx}

\ifCLASSOPTIONcompsoc  \usepackage[caption=false,font=normalsize,labelfont=sf,textfont=sf]{subfig}
\else
  \usepackage[caption=false,font=footnotesize]{subfig}
\fi

\usepackage{float}
\usepackage{stfloats}
\usepackage{url}

\usepackage[colorlinks]{hyperref}
\usepackage[colorinlistoftodos]{todonotes}
\usepackage{verbatim}

\newtheorem{Ex}{Example}[]
\newtheorem{Def}{Definition}[]
\newtheorem{Prop}{Proposition}[]
\newtheorem{Teo}{Theorem}[]

\newtheorem{Obs}{Remark}[]
\newtheorem{Lema}{Lemma}[]

\newcommand{\Z}{\mathbb{Z}}
\newcommand{\Q}{\mathbb{Q}}
\newcommand{\C}{\mathbb{C}}
\newcommand{\N}{\mathbb{N}}

\newcommand{\R}{\mathbb{R}}

\newcommand{\h}{\mathbb{H}}
\newcommand{\F}{\mathbb{K}}
\newcommand{\E}{\mathbb{E}}
\newcommand{\la}{\mathbb{L}}
\newcommand{\Partopo}{\mathcal{P}}

\newcommand{\vor}{\mathscr{V}}

\newcommand{\cod}{\mathcal{C}}
\newcommand{\ring}{\mathcal{R}}

\newcommand{\ds}{\displaystyle}

\newcommand{\ii}{i}
\newcommand{\jj}{j}
\newcommand{\kk}{k}
\newcommand{\ww}{\omega}

\newcommand{\Lip}{\mathscr{L}}

\newcommand{\Hu}{\mathscr{H}}

\newcommand{\primo}{\mathfrak{P}}
\newcommand{\cqd}{\hfill $\Box$}

\newcommand{\fundR}{\mathscr{R}}

\begin{document}

%
% paper title
% Titles are generally capitalized except for words such as a, an, and, as,
% at, but, by, for, in, nor, of, on, or, the, to and up, which are usually
% not capitalized unless they are the first or last word of the title.
% Linebreaks \\ can be used within to get better formatting as desired.
% Do not put math or special symbols in the title.
\title{Multilevel lattice codes from Hurwitz quaternion integers}
%
%
% author names and IEEE memberships
% note positions of commas and nonbreaking spaces ( ~ ) LaTeX will not break
% a structure at a ~ so this keeps an author's name from being broken across
% two lines.
% use \thanks{} to gain access to the first footnote area
% a separate \thanks must be used for each paragraph as LaTeX2e's \thanks
% was not built to handle multiple paragraphs
%

\author{Juliana~G. F. Souza,~\IEEEmembership{Student Member, ~IEEE,}
        Sueli~I. R. Costa,~\IEEEmembership{Member,~IEEE,}
        and~Cong~Ling,~\IEEEmembership{Member,~IEEE}% <-this % stops a space
\thanks{Manuscript received Month XX, 2024. An earlier version of parts of this work was submmited at the 2024 IEEE International Symposium on Information Theory. This work is partially supported by Brazilian foundations Coordination for the Improvement of Higher Education Personnel (CAPES -- Financial Code 001), FAPESP (2020/09838 -- 0).}%
\thanks{Juliana G. F. Souza is with the Institute of Mathematics, Statistics and Scientific Computing (IMECC), University of Campinas (Unicamp), Campinas, São Paulo 13083-859, Brazil (e-mail: julianagfs@ime.unicamp.br).}% <-this % stops a space
\thanks{Sueli I. R. Costa is with the Institute of
Mathematics, Statistics and Scientific Computing (IMECC), University of Campinas (Unicamp), Campinas, São Paulo 13083-859, Brazil (e-mail: sueli@unicamp.br).}% <-this % stops a space
\thanks{Cong Ling is with the Department of Electrical and Electronic Engineering,
Imperial College London, SW7 2AZ London, U.K. (e-mail: cling@ieee.org).}}

% The paper headers
%\markboth{Journal of \LaTeX\ Class Files,~Vol.~14, No.~8, August~2015}%
%{Shell \MakeLowercase{\textit{et al.}}: Bare Demo of IEEEtran.cls for IEEE Journals}

\maketitle

% As a general rule, do not put math, special symbols or citations
% in the abstract or keywords.
\begin{abstract}
This work presents an extension of the Construction $\pi_A$ lattices proposed by Huang and Narayanan, to Hurwitz quaternion integers. This construction is provided by using an isomorphism from a version of the Chinese remainder theorem applied to maximal orders, in contrast to natural orders in prior works. Exploiting this map, we analyse the performance of the resulting multilevel lattice codes and highlight via computer simulations their notably reduced computational complexity provided by the multistage decoding. Moreover, it is shown that there is a sequence of Construction $\pi_A$ lattices that attain with high probability the Poltyrev-limit.
\end{abstract}

% Note that keywords are not normally used for peerreview papers.

% For peer review papers, you can put extra information on the cover
% page as needed:
% \ifCLASSOPTIONpeerreview
% \begin{center} \bfseries EDICS Category: 3-BBND \end{center}
% \fi
%
% For peerreview papers, this IEEEtran command inserts a page break and
% creates the second title. It will be ignored for other modes.
\IEEEpeerreviewmaketitle

\section{Introduction}
% The very first letter is a 2 line initial drop letter followed
% by the rest of the first word in caps.
% 
% form to use if the first word consists of a single letter:
% \IEEEPARstart{A}{demo} file is ....
% 
% form to use if you need the single drop letter followed by
% normal text (unknown if ever used by the IEEE):
% \IEEEPARstart{A}{}demo file is ....
% 
% Some journals put the first two words in caps:
% \IEEEPARstart{T}{his demo} file is ....
% 
% Here we have the typical use of a "T" for an initial drop letter
% and "HIS" in caps to complete the first word.
\IEEEPARstart{L}{attices} have been used in communication scenarios for several decades, regarding coding for reliable and secure transmission through different channels. Some coding techniques employed rely on a randomized collection of nested lattice codes derived from Construction A, which can achieve the capacity of the additive white Gaussian noise (AWGN) channel \cite{erez2004achieving,erez2005lattices}. However, decoding Construction A lattices typically involves decoding the underlying linear codes over a finite ring, leading to substantial computational complexity in decoding.

To address this challenge, Huang and Narayanan proposed in \cite{huang2017construction} a multilevel lattice construction known as Construction $\pi_A$ firstly considering integers $\Z\subset\R$, Gaussian integers $\Z[\ii]\subset\C$, and Eisenstein integers $\Z[\ww]\subset\C$. This construction can be viewed as a special case of Construction A, applied to codes represented as the Cartesian product of $r\in \N$ linear codes. These codes are considered over the residue class rings $R/p_1R,\ldots,R/p_kR$, where $R=\Z,\Z[\ii]$ or $\Z[\ww]$ and $p_i$ is a prime number in the related ring. Such constructions are made possible through a ring isomorphism between the product of finite rings and the quotient rings $\left(R/\prod_{j=1}^{r}p_j R\right)$, established by the Chinese Remainder Theorem (CRT). Additionally, due to its multilevel nature, Construction $\pi_A$ lattices offer the advantage of multistage decoding, where each class representative is decoded level by level. Further extensions of this construction were proposed in \cite{huang2018lattices} using imaginary quadratic fields and in \cite{huang2018layered} for natural orders over cyclic division algebras. In \cite{huang2017golden}, it is proposed a scheme using partitions of the golden code, where it can be viewed as a space-time (ST) code derived from a cyclic division algebra. The authors addressed the difficulty of applying the CRT in non-commutative settings by embedding the cyclic division algebra into a commutative structure, enabling partitions of the golden code.

In the context of communication theory, cyclic division algebras, particularly through the use of natural and maximal orders, have been proved useful in the design of ST codes. Several studies have explored natural orders for explicit constructions of fully diverse and fast-decodable ST codes \cite{oggier2006perfect, elia2006explicit, alamouti1998}. Later, the use of maximal orders has emerged as an effective strategy, enabling denser constructions without compromising crucial parameters like minimum distance. The initial work on maximal orders to design ST codes was proposed in \cite{hollanti2008maximal}, subsequently leading to the derivation of the densest ST codes \cite{vehkalahti2009densest}. Furthermore, various studies have explored the numerous advantages of employing maximal orders in quaternionic settings \cite{cintya2020maximal, alves2015lattices}. These orders offer significant advantages, particularly in multiple-input/single-output (MISO) \cite{hollanti2008maximal} and multiple-input/multiple-output (MIMO) transmissions \cite{hollanti2009asymetric, vehkalahti2009densest}.

A particularly rich structure arises from the Hurwitz quaternion integers $\Hu$, considered here, which forms a maximal order in the Hamilton quaternion algebra $\h$. The quotient rings of the format $\Hu/p\Hu$ (p odd prime) provide code constructions with favourable properties. An important relationship regarding applications is the fact that $\Hu$ is a quaternion-valued representation of the $D_4^*$ lattice,  providing the best lattice sphere packing efficiency in this dimension  \cite{Con2013, conway2003quaternions}. For instance, the \textit{Hurwitz lattice} (the lattice associated to $\Hu$) has the double of the packing density of the lattice $\Z^4$ which is associated to the natural order of the Lipschitz quaternion integers. These properties have motivated its use in optical communications \cite{agrell2009, karlsson2011, zetterberg1977, welti1974}, broadcast transmission over the AWGN channel \cite{natarajan2015lattice,huang2018layered} and MIMO transmissions \cite{freudenberger2017generalized,stern2018quaternion,ugrelidze2020new, stern2022algorithms}. Notably, with a more theoretical approach, it was shown that Hurwitz lattice in dimension $4n$ improve the Minkowski-Hlawka lower bound \cite{vance2011improved}, highlighting their strong potential in lattice coding.

In this paper, we propose a novel extension of Construction $\pi_A$ lattices \cite{huang2017construction} utilizing the maximal order of Hurwitz quaternion integers and a modified version of the CRT adapted to this non-commutative structure. This approach enables the construction of multilevel codes with different sizes than those obtained in \cite{huang2018layered} and supports efficient multistage decoding. We adopt Loeliger's averaging approach \cite{loeliger1997averaging} instead of Forney's \cite{forney2000sphere} as in \cite{huang2017construction} to demonstrate that the resulting lattice achieves the Poltyrev-limit over the AWGN channel, and to analyse their computational complexity via numerical simulations. We also include a preliminary investigation of Hurwitz-based Construction $\pi_A$ lattice codes in the context of index coding problems, where their algebraic structure may offer advantages in side information exploitation.

The remaining of this paper is organized as follows: Section II provides the necessary background on lattice codes and quaternion algebras. Section III presents the extension of Construction $\pi_A$ to Hurwitz integers. Section IV describes a decoding algorithm for Construction $\pi_A$ lattices over the Hurwitz quaternion integers and discusses its decoding complexity. In Section V it is shown that this extension can achieve the Poltyrev-limit for channel coding. Section VI presents an example of the use of this setting to index coding problem, and in Section VII conclusions and future perspectives are drawn.

\subsection{Notations}

We denote by $\Z, \Z[\ii]=\{a+b\ii \ | \ a,b\in\Z\}$ and $\Z[\ww]=\{a+b\ww \ | \ a,b\in\Z, w=\frac{-1+\sqrt{3}\ii}{2}\}$, the ring of integers, the ring of Gaussian integers and the ring of Eisenstein integers, respectively. Also, as usual $\N, \ \Q, \ \R, \ \C$ and $\h$ represent the set of natural numbers, the field of rational numbers, the field of real numbers, the field of complex numbers and the skew-field of Hamilton quaternions, respectively. An arbitrary field is denoted by $\F$ and  $p\in\N$ is called a rational prime if $p>1$ is a prime number. A code with cardinality $M$ is denoted as $|\cod| = M$. The modulo reduction is denoted as $a\mod q = a - q \cdot \lfloor q^{-1}\cdot a \rceil$, where $\lfloor \ \cdot \ \rceil$ is the nearest integer in the considered ring. To turn more clear, it is also mentioned the ring being considered. For example, $a\mod q\Hu$ is $a\mod q$ in the ring of Hurwitz quaternion integers.

\section{Preliminaries}

\subsection{Lattices and lattice codes}

We summarize next some concepts and properties related to lattices \cite{Con2013}, \cite{Sue2018}.

	A \textit{lattice} $\Lambda$ can be defined as an additive discrete subgroup of $\R^n$. Equivalently, a lattice is a subset of $\R^n$ generated by all integer linear combinations of $m$ independent vectors, $v_1,\ldots,v_m\in\R^n$. A matrix $B$ whose columns are the vectors $v_i$ is called a \textit{generator matrix} of $\Lambda$. We deal here only with full-rank lattices ($m=n$).

	A set $\fundR\subset\R^n$ is a \textit{fundamental region} of $\Lambda$, iff
	\begin{enumerate}
		\item[(i)] For $x,y\in\Lambda$, where $x\neq y$, $(x+\fundR)$ and $(y+\fundR)$ intersect at most on their boundaries and,
		\item[(ii)] $\ds\bigcup_{x\in\Lambda} (x+\fundR)=\R^n$, i.e., the entire $\R^n$ space is tiled  through translations by points of $\Lambda$.
	\end{enumerate}

An example of fundamental region of $\Lambda$ is the \textit{fundamental parallelotope}, $\Partopo(B)$ associated to a generator matrix $B$ and defined as,
\[
\Partopo(B)=\left\{ \sum_{i=1}^n \alpha_i v_i, 0\leq\alpha_i<1, \alpha_i\in\R \right\}.
\]

The \textit{volume} of a lattice $\Lambda$, $\text{vol}(\Lambda)=|\det(B)|$, is the volume of any fundamental region of $\Lambda$.

Another fundamental region associated to a lattice $\Lambda$ is the \textit{Voronoi region}, $\vor_\Lambda(x)$, at a point $x\in\Lambda$:
	\begin{equation}
		\vor_\Lambda(x)=\{y\in\R^n;|x-y|\leq|\lambda-y|, \ \forall \lambda\in\Lambda\}.
	\end{equation}

Given a point $z\in\R^n$ and a lattice $\Lambda\subset\R^n$, we define $Q_{\Lambda}(z)$ as the \textit{closest lattice point} to $z$ as
	\begin{equation}
		Q_{\Lambda}(z) = \lambda\in\Lambda; \ \ ||z-\lambda||\leq||z-\hat{\lambda}|| \ \ \ \ \forall \ \hat{\lambda}\in\Lambda,
	\end{equation}
\noindent where ties are broken systematically as in \cite{stern2022algorithms}, for example. Subtracting the closest lattice point $Q_{\Lambda}(z)$ from $z$ wraps the real vector $z$ into the Voronoi region at the origin, $\vor_{\Lambda}(0)=\vor_\Lambda$. This operation, called the \textit{modulo-$\Lambda$} operation, is denoted as

	\begin{equation}
		z\mod\Lambda = z - Q_{\Lambda}(z).
	\end{equation}

The \textit{minimum distance} between two distinct points of a lattice $\Lambda$ corresponds to the minimum Euclidean norm of non-zero vectors in $\Lambda$, i.e.,
	\begin{equation}
		d_{\text{min}}(\Lambda)=\ds\min_{0\neq x\in\Lambda}\|x\|.
	\end{equation}

The \textit{sphere packing} of a lattice $\Lambda$ is the union of all the translated balls with radius $\rho=d_{\text{min}}(\Lambda)/2$ centred at points of $\Lambda$ and $\rho$ is called the \textit{packing radius} of $\Lambda$.

A \textit{linear code} over $\Z/q\Z = \Z_q$, the ring of integers modulo $q$, is a subset $\cod\subset\Z_q^n$ closed under addition. 

A method for constructing lattices from linear codes is the well known \textit{Construction A}. A Construction A lattice is usually defined from linear codes over $\Z_q$, but can also be considered from codes over $\Z[\ii]$ and $\Z[\ww]$, number fields or division algebras \cite{Con2013, kositwattanarerk2015construction, vehkalahti2014constructions}.

\begin{Def}[Construction A]
	\label{def1}
	Let $q>1$ be an integer. Let $k, n\in\N$ be integers such that $k\leq n$, and let $G$ be an $n\times k$ generator matrix of a linear code over $\Z_q$. Construction A consists of the following steps:
	\begin{enumerate}
		\item Consider the linear code $\cod=\{x=G\cdot y: y\in \Z_q^{k}\}$, where all operations are over $\Z_q$.
		\item ``Expand'' $\cod$ to a lattice in $\Z^n$ defined as:
		\begin{equation*}
			\Lambda_A(\cod)=\{x\in\Z^n: x\mod q\in\cod\}=\cod+q\Z^n.
		\end{equation*}
	\end{enumerate}
\end{Def}

It is shown that $\Lambda_A(\cod)$ is a full-rank lattice, $q\Z^n\subset\Lambda_A(\cod)\subset\Z^n$, and that the volume of this lattice is $q^n/M$, where $M$ is the size of the code $\cod$ \cite{Con2013}.

In \cite{loeliger1997averaging}, an important result of the Minkowski-Hlawka theorem \cite{hlawka1943geometrie, Cas1997} was used to show that by randomly selecting a code from the set of all $(n,k)$-linear code over $\Z_p$, $p$ prime, Construction A produces lattices that are suitable for channel coding as $p$ tends to infinity.

Consider the unconstrained AWGN channel $y=x+z$, where $y$ is the received signal, $x\in\Lambda$ is the transmitted signal with no power constraint, and each element in $z$ is i.i.d. $\sim \mathcal{N}(0,\sigma^2)$. Consider a target error probability $0<\epsilon<1$, the \textit{normalized volume to noise ratio} (NVNR) of a lattice $\Lambda\subset\R^n$ is defined as
\begin{equation*}
    \mu(\Lambda,\epsilon)=\dfrac{\text{vol}(\Lambda)^{2/n}}{\sigma^2(\epsilon)},
\end{equation*}

\noindent where $\sigma^2(\epsilon)$ is the value of $\sigma^2$ such that $P_e(\Lambda,\sigma^2)$ is equal to $\epsilon$, for some error probability $0<\epsilon<1$.

	Then, a sequence of lattices $\Lambda_n$ of increasing dimension is \textit{Poltyrev-good} \cite{zamir2014lattice} if
	\begin{align}
		\lim_{n\rightarrow\infty} \dfrac{\text{vol}(\Lambda)^{2/n}}{\sigma^2(\epsilon)} = 2\pi e \ \text{for all} \ 0<P_e<1.
      \label{poltyrev_limit}
	\end{align}

In \cite{poltyrev1994coding}, this setup was proposed for the unconstrained AWGN channel case and (\ref{poltyrev_limit}) is known as the \textit{Poltyrev limit}.

\subsection{Quaternion Algebras}

Let $\F$ be a field. An \textit{algebra} $B$ over $\F$ (or $\F-$algebra) is a vector space over $\F$
equipped with an associative multiplication
\begin{align*}
    B\times B&\rightarrow B\\
    (x,y)&\mapsto xy,
\end{align*}

\noindent satisfying $(k_1 x+k_2 y)z = k_1(xz)+k_2(yz) \ \text{and} \ x(k_1y + k_2z)=k_1(xy)+k_2(xz)$ for all $k_1, k_2\in \F$, $x,y,z\in B$. %Thus an $\F-$algebra is a ring which is also a vector space and multiplication is bilinear.

\begin{Def}[Quaternion algebra]
	Let $\F$ be a field with characteristic different from 2. A quaternion algebra over $\F$ is an $\F$-algebra admitting a basis of four elements, denoted $1, \ii, \jj, \kk$, which satisfy the following relations: $1$ is the neutral element for multiplication, and
	$$\ii^2 = a, \ \ \jj^2=b \ \ \ii\jj=\kk=-\jj\ii,$$
	
	\noindent for some non-vanishing elements $a,b\in\F$. We denote this algebra by $(a,b)_{\F}$.
\end{Def}

The classical example to be considered here is the quaternion algebra over the reals due to Hamilton \cite[(1848)]{hamilton1848},
$$\h = (-1,-1)_{\R} = \{a_0 + a_1\ii + a_2\jj + a_3\kk: (a_0, a_1, a_2, a_3)\in\R\}.$$

If $B$ is assumed to be a quaternion algebra over the field $\F=\Q$ of rational numbers with $a,b<0$, $B=(a,b)_{\Q}$, $B$ is also called a \textit{definite quaternion algebra}.

We define a \textit{full} $\Z-$\textit{lattice} $\Gamma$ in a finite-dimensional definite quaternion algebra $B$ as a finitely generated $\Z-$submodule $\Gamma\subset B$ that contains a $\Q-$basis of $B$. 

\begin{Def}[Order]
	An order $O\subset B = (a,b)_{\Q}, a,b<0$ is a full $\Z$-lattice that is also a subring having $1\in B$.
\end{Def}

\begin{Ex}
Consider $B=(a,b)_{\Q}$ a definite quaternion algebra. Then
	$$O = \Z + \Z\ii + \Z\jj + \Z\kk,$$
is an order in $B$, called a \textit{natural order}.
\end{Ex}

Due to the non-commutativity of multiplication, it is necessary to distinguish between left ideals, right ideals and two-sided ideals. Throughout this paper, we focus only on left operations, although equivalent results can be obtained for right operations as well.

\begin{Def}[Left ideal]
A subset $I$ of an order $O$ of a definite quaternion algebra $B=(a,b)_{\Q}$ is a left-ideal if it satisfies the following conditions:
\begin{enumerate}
	\item $0\in I$;
	\item $x+y\in I$ if $x,y\in I$;
	\item $\lambda x\in I$ for any $\lambda\in O$ and $x\in I$.
\end{enumerate}
\end{Def}

\noindent \textit{Right ideals} are defined analogously, and a \textit{two-sided ideal} is a subset that is both a left and a right ideal. 

\begin{Def}[Left prime ideal, \cite{reyes2010one}]
	We define $\mathfrak{p} \subsetneq O$ a \textit{left prime} ideal whenever $\mathfrak{a},\mathfrak{b} \subset O$ are left ideals such that $\mathfrak{a}\mathfrak{b} \subset \mathfrak{p}$ implies that either $\mathfrak{a} \subset \mathfrak{p}$ or $\mathfrak{b} \subset \mathfrak{p}$.
\end{Def}

\begin{Def}[Prime ideal, \cite{reiner1975maximal}]
	A two-sided ideal $\mathfrak{P} \subsetneq O\subset B$ is said to be a \textit{prime ideal} if, for two-sided ideals  $\mathfrak{U}, \mathfrak{W} \subset O$, we have, 	
	$$\mathfrak{U}\cdot\mathfrak{W}\subset\mathfrak{P} \Rightarrow \mathfrak{U}\subset\mathfrak{P} \  \ \textrm{or} \ \ \mathfrak{W}\subset\mathfrak{P}.$$
\end{Def}

\begin{Def}[Maximal Order]
	An order $O\subset B=(a,b)_{\Q}, a,b<0$ is maximal if it is not properly contained in another order of $B$.
\end{Def}

\begin{Ex}
	\label{hurwitz}
	Let  $B=(-1,-1)_{\Q}$ and
	\begin{equation}
	    \Lip=\{a_1 1+ a_2\ii + a_3\jj + a_4\kk | a_1,\ldots,a_4\in\Z\},
	\end{equation}	
	\noindent where $(1, \ii, \jj, \kk)$ is the standard basis of $B$. A direct verification shows that $\Lip$ is an order of $B$ called the \textit{Lipschitz order}. This is not a maximal order \cite{voight2021quaternion}.
\end{Ex}

The maximal order in $B=(-1,-1)_{\Q}$ is given by the Hurwitz quaternion integers \cite{voight2021quaternion}, defined next.

\begin{Def}[Hurwitz quaternion integers]
The Hurwitz quaternion integers (Hurwitz integers), $\Hu\subset\h$, is defined as the set
	$$\Hu = \left\{a_1+a_2\ii+a_3\jj+a_4\left(\frac{1+\ii+\jj+\kk}{2}\right) \ | \ a_1,\ldots, a_4\in\Z\right\}.$$
\end{Def}
 
From now on, we restrict our work to $B=(-1,-1)_{\Q}$ and the maximal order $\Hu$ of the Hurwitz integers. Note that $\Hu$ is composed by all quaternions having each component as an integer or a ``half-integer''.

\begin{Ex}
	$1+\ii\in\Hu$, $\frac{3}{2}-\frac{\ii}{2}+\frac{5\jj}{2}-\frac{9\kk}{2}\in\Hu$, but $1+\frac{\ii}{2}\notin\Hu$.
\end{Ex}

For an element $\alpha= a+b\ii+c\jj+d\kk \in \Hu$, its conjugated $\overline{\alpha}$ is $a-b\ii-c\jj-d\kk$. The \textit{norm} of $\alpha$ is $\text{Nrm}(\alpha)=\alpha \cdot \overline{\alpha} = \overline{\alpha}\cdot\alpha = a^2 +b^2 +c^2 +d^2\in \Z$ and $Re(\alpha) = a$ represents the real part of $\alpha$. An element $\varepsilon$ is a unit iff $\text{Nrm}(\varepsilon)=1$. The set of units, $\Hu^{\times}$, is composed of $24$ units, namely
\[\Hu^{\times} = \left\{\pm 1, \pm\ii, \pm\jj, \pm\kk, \dfrac{\pm1\pm\ii\pm\jj\pm\kk}{2}\right\}.\]

The set of Hurwitz integers has some important properties, such as admitting a left Euclidean algorithm which can be obtained as in the commutative case \cite{voight2021quaternion}.

Using left division, we can define the left greatest common divisor, and also a version of Bézout's identity holds.

\begin{Def}[Left division]
	\label{leftdiv}
	Let $\alpha, \beta\in O$. We say $\beta$ left divides $\alpha$ (or $\alpha$ is a right multiple of $\beta$) and write $\beta\rceil\alpha$ if there exists $\gamma\in O$ such that $\alpha = \beta\gamma$.
\end{Def}

\begin{Def}[Left Greatest Common Divisor (LGCD)]
	Let $\alpha, \beta\in \Hu$. We say that $\delta\in\Hu$ is a left greatest common divisor of $\alpha$ and $\beta$ if:
	\begin{itemize}
		\item $\delta$ is a left divisor of both $\alpha$ and $\beta$;
		\item if $\delta'\in\Hu$ is a left divisor of both $\alpha$ and $\beta$, then $\delta'$ is a left divisor of $\delta$.
	\end{itemize}
\end{Def}

\begin{Prop}[Bézout's theorem, \cite{voight2021quaternion}]
	For all $\alpha, \beta\in \Hu$ not both zero, there exist $\mu, \nu\in\Hu$ such that $\mu\alpha+\nu\beta = \delta$, where $\delta$ is a left greatest common divisor of $\alpha$ and $\beta$.
\end{Prop}

\begin{Def}[Irreducible] An element $\pi\in\Hu$ is irreducible if whenever $\pi=\alpha\beta$ with $\alpha,\beta\in\Hu$ then either $\alpha\in\Hu^{\times}$ or $\beta\in\Hu^{\times}$.
\end{Def}

Note that in $\Z[\ii]$ and $\Z[\ww]$ the concept of irreducible corresponds to ``prime'' as in the natural numbers. But in $\Hu$ we may have an irreducible quaternion $\pi$ to divide from the left a product $\alpha\beta$, $\pi\rceil \alpha\beta$, without having either $\pi\rceil \alpha$ or $\pi\rceil \beta$. For example, let $\delta=(1+\jj+\kk)(1+\jj) = -\ii + 2\jj+\kk$ and $\pi=1+\ii-\jj$, then $\pi\rceil \delta$ since we can write $\delta=(1+\jj)\pi$ but we do not have either $\pi\rceil (1+\jj+\kk)$ or $\pi\rceil (1+\jj)$.

For Hurwitz integers, it is shown that the irreducible elements are those with norm equal to a rational prime.

\begin{Prop}[\cite{voight2021quaternion}, p.171]
	Let $\pi\in\Hu$. Then $\pi$ is irreducible if and only if $\textrm{Nrm}(\pi)=p\in\Z$ is a rational prime.
\end{Prop}

Other remarkable properties of Hurwitz integers, such as ``unique factorization'', have been discussed in \cite{voight2021quaternion} and \cite{conway2003quaternions}.

A result that provides the construction of multilevel lattice codes is the CRT. Inspired by the Construction $\pi_A$ proposed in \cite{huang2017construction, huang2018lattices} and \cite{huang2018layered}, we found the following version of CRT that can be applied to maximal orders over definite quaternion algebras and therefore to the Hurwitz integers.

\begin{Teo}[Chinese remainder theorem for maximal orders, (\cite{reiner1975maximal}, Chap. 6)]
	\label{crtquaternio}
	Let $O$ be a maximal order, and let $\primo_1,\ldots,\primo_n \subsetneq O$ be distinct (two-sided) prime ideals.
	Let $\mathfrak{P} = \prod_{i=1}^{n} \primo_i^{a_i}$, $a_i \in \Z$, $i=1,\ldots,n$. If $a_i \geq 0$ for all $i=1,\ldots,n$, then there is a ring isomorphism
	$$O/\mathfrak{P}\cong (O/\primo_1^{a_1})\times \cdots \times (O/\primo_n^{a_n}).$$
\end{Teo}

The following result together with the previous \textit{Theorem \ref{crtquaternio}} will be used in the proof of \textit{Theorem \ref{isomorfismH}} in the next section which extends Construction $\pi_A$ to Hurwitz integers.

As it is remarked in \cite{natarajan2015lattice}, given an odd rational prime $p$ there exists an irreducible quaternion $\pi$ with $\text{Nrm}(\pi)=p$ such that $Re(\pi)\in\{1,2\}$. This is a consequence of three-square's Legendre theorem \cite{ireland1990classical}.

\begin{Teo}
	\label{crt2}
	Let $\Hu \subset B = (-1,-1)_{\Q}$, and let $p\Hu$ be the prime ideal generated by an odd rational prime $p \in \Z$.
	%such that $p = \pi \cdot \overline{\pi} = \textrm{Nrm}(\pi)$ for some irreducible element $\pi \in O$
	Then there exists an isomorphism,
	\begin{align*}
	    \Psi: \Hu/p\Hu &\rightarrow \Hu/\Hu\pi \times \Hu/\Hu\overline{\pi},
	\end{align*}
	
\noindent where $\Hu\pi$ and $\Hu\overline{\pi}$ are left prime ideals with $\text{Nrm}(\pi)=p$ and $Re(\pi)\in\{1,2\}$.
\end{Teo}

\noindent \textit{Proof:} Let $p=\pi\overline{\pi}=\overline{\pi}\pi$ with $Re(\pi)\in\{1,2\}$. Consider the map
\begin{align*}
	    \Psi: \Hu/p\Hu &\rightarrow \Hu/\Hu\pi \times \Hu/\Hu\overline{\pi}\\
      a + p\Hu &\mapsto (a + \Hu\pi, a + \Hu\overline{\pi}).
\end{align*}

It is straightforward to see that this map is linear. Since $p\Hu$ is a prime ideal and $p\Hu=(\Hu\overline{\pi})\pi=(\Hu\pi)\overline{\pi}$, we have
{\small \begin{align*}
    \Psi(\alpha+\beta+p\Hu)= &((\alpha + \beta + p\Hu)+\Hu\pi, (\alpha + \beta + p\Hu)+\Hu\overline{\pi})\\
    = &(\alpha + \beta + \Hu\pi, \alpha + \beta + \Hu\overline{\pi})\\
    = &(\alpha + \Hu\pi, \alpha + \Hu\overline{\pi}) + (\beta + \Hu\pi, \beta + \Hu\overline{\pi})\\
%   = &((\alpha + p\Hu) + \Hu\pi, (\alpha + p\Hu) + \Hu\overline{\pi}) + \\
%      & +((\beta + p\Hu) + \Hu\pi, (\beta + p\Hu) + \Hu\overline{\pi})\\
    = & \Psi(\alpha + p\Hu) + \Psi(\beta + p\Hu)  .
\end{align*}}

Let's show now that $\Psi$ is surjective. In fact, given in $(\alpha+\Hu\pi, \beta +\Hu\overline{\pi})$, since $LGCD(\pi,\overline{\pi})=1$, by Bèzout's theorem, there exist $\gamma_1,\gamma_2\in\Hu$ such that,
\begin{equation}
    \gamma_1\overline{\pi}+\gamma_2\pi = 1 + p\Hu.
\end{equation}

If $Re(\pi)=1$, we can take $\gamma_1=\gamma_2=\gamma=\frac{p+1}{2}\in\Z$ and, if $Re(\pi)=2$ we can take $\gamma_1=\gamma_2=\gamma=\frac{p+3}{4}\in\Z$. Then
\begin{equation}
    \gamma\overline{\pi}+\gamma\pi = 1 + p\Hu, \tag{*}
\end{equation}

\noindent for both cases.

Now consider $\theta\in\Hu/p\Hu$ such that
\[
\theta = (\alpha\gamma\overline{\pi}+\beta\gamma\pi) + p\Hu.
\]

Then,
{\small \begin{align*}
    \Psi(\theta) = & (\theta + \Hu\pi, \theta + \Hu\overline{\pi})\\
     = & ((\alpha\gamma\overline{\pi}+\beta\gamma\pi) + p\Hu + \Hu\pi, (\alpha\gamma\overline{\pi}+\beta\gamma\pi) + p\Hu + \Hu\overline{\pi})\\
     = & ((\alpha\gamma\overline{\pi}+\beta\gamma\pi) + \Hu\pi, (\alpha\gamma\overline{\pi}+\beta\gamma\pi) + \Hu\overline{\pi})\\
     = & (\alpha\gamma\overline{\pi} + \Hu\pi, \beta\gamma\pi + \Hu\overline{\pi}).
\end{align*}}

From $(*)$, considering the classes in $\Hu\pi$ and $\Hu\overline{\pi}$ we have
\begin{align*}
    \gamma\overline{\pi} + \Hu\pi = 1 + \Hu\pi & \Rightarrow \gamma\overline{\pi} = 1 + \Hu\pi,\\
    \gamma\pi + \Hu\overline{\pi} = 1 + \Hu\overline{\pi} & \Rightarrow \gamma\pi = 1 + \Hu\overline{\pi}.
\end{align*}

Then,
\begin{align*}
    \Psi(\theta) = & (\alpha\gamma\overline{\pi} + \Hu\pi, \beta\gamma\pi + \Hu\overline{\pi})\\
     = & (\alpha + \Hu\pi, \beta + \Hu\overline{\pi}).
\end{align*}

To conclude that $\Psi$ is an isomorphism, we use the fact that for any $\alpha\in\Hu$, the number of classes of $\Hu/\Hu\alpha$, denoted by $|\Hu/\Hu\alpha|$ is $\text{Nrm}(\alpha)^2$, \cite{conway2003quaternions}. Therefore, the domain and the image of $\Psi$ have the same number of elements,
\begin{align*}
    \ds\left|\dfrac{\Hu}{\Hu p}\right|=&(\text{Nrm}(p))^2=(p^2)^2 \ \text{and} \\
    \ds\left|\dfrac{\Hu}{\Hu \pi}\right|\cdot\left|\dfrac{\Hu}{\Hu \overline{\pi}}\right|= & (\text{Nrm}(\pi))^2 \cdot (\text{Nrm}(\overline{\pi}))^2 = p^2\cdot p^2,
\end{align*}

\cqd

Note also that, under the notation in the proof of the last theorem for any $(\alpha,\beta)\in \Hu/\Hu\pi \times \Hu/\Hu\overline{\pi}$,
\begin{equation}
    \Psi^{-1}(\alpha,\beta)=\gamma(\alpha\overline{\pi}+\beta\pi) + p\Hu.
    \label{eqcrt2}
\end{equation}

\section{Multilevel lattice codes over Hurwitz integers}

Inspired in \cite{huang2017construction}, we propose an extension of Construction $\pi_A$ to the Hurwitz integers. As seen, the Hurwitz integers have good number-theoretical behaviour. Therefore, it is natural to investigate this extension by considering their applications in communications. Another nice aspect of the Hurwitz integers is their close relationship with the dual lattice $D_4^*$. In particular, the basis $\{1,\ii,\jj,\frac{1+\ii+\jj+\kk}{2}\}$ for $\Hu$ corresponds to the generator matrix $B_{D_4^{*}}=\{(1,0,0,0),(0,1,0,0),(0,0,1,0),(1/2,1/2,1/2,1/2)\}$, \cite{Con2013}.

A strong motivation for extending Construction $\pi_A$ to Hurwitz integers was also the paper \cite{natarajan2015lattice}. There, it is proposed the construction of lattice codes over $\Z, \Z[\ii], \Z[\ww]$ and $\Hu$, obtaining the so-called index lattice codes \cite{natarajan2015lattice2, huang2017lattice}. In \cite{natarajan2015lattice}, it is used lattice constellations along with an injective map which is associated to the CRT in the commutative cases.

For our construction, we will present an isomorphism that will serve as a crucial component to obtain lattices over Hurwitz integers.

\begin{Teo}
\label{isomorfismH}
    Let $p_1, p_2,\ldots,p_r$ be distinct odd rational primes and let $q=\prod_{j=1}^r p_j$. Consider $\pi_j\in\Hu$ such that $\text{Nrm}(\pi_j)=p_j$, for $j=1,\ldots,r$. There exists a ring isomorphism 
    \[\Hu/q\Hu \cong \Hu/\Hu\pi_1\times\Hu/\Hu\overline{\pi}_1\times\ldots\times \Hu/\Hu\pi_k\times\Hu/\Hu\overline{\pi}_r.\]
\end{Teo}

\noindent \textit{Proof:} We have
\begin{align*}
	\Hu/q\Hu &\stackrel{(a)}{\cong} \Hu/p_1\Hu \times \dots \times \Hu/p_r\Hu\\
	&\stackrel{(b)}{\cong} \Hu/\Hu\pi_1 \times \Hu/\Hu\overline{\pi}_1 \times \dots \times \Hu/\Hu\pi_k \times \Hu/\Hu\overline{\pi}_r,
\end{align*}

\noindent where (a) follows from \textit{Theorem \ref{crtquaternio}} and (b) from \textit{Theorem \ref{crt2}}. Therefore, there exists a ring isomorphism $\varphi$ between $\Hu/q\Hu$ and the product $\Hu/\Hu\pi_1 \times \Hu/\Hu\overline{\pi}_1 \times \dots \times \Hu/\Hu\pi_k \times \Hu/\Hu\overline{\pi}_r$.

\cqd

Our objective is to determine an explicit expression for such an isomorphism in order to apply it in the Construction $\pi_A$. 

Let's consider distinct odd rational primes $p_1, \dots, p_r$, $q = \prod_{i=1}^r p_i$, $q\Hu$, and let $p_i\Hu$ be two-sided ideals generated by $q$ and $p_i$, respectively. Let $\Hu\pi_i$ and $\Hu\overline{\pi}_i$ be prime left ideals generated by the irreducible elements $\pi_i, \overline{\pi}_i$ of $\Hu$, $p_i=\pi_i\cdot\overline{\pi}_i$. The steps to obtain an ring isomorphism $\varphi: \Hu/q\Hu \rightarrow \Hu/\Hu\pi_1 \times \Hu/\Hu\overline{\pi}_1 \times \dots \times \Hu/\Hu\pi_k \times \Hu/\Hu\overline{\pi}_r$ are as follows.

From (\ref{eqcrt2}), we have
    \begin{align*}
            \Psi^{-1}_{i}(a_i^{(1)}, a_i^{(2)}) = \gamma_i(\overline{\pi}_i a_i^{(1)} + \pi_i a_i^{(2)}) \mod p_i\Hu,
    \end{align*}
\noindent where $\gamma_i\in\Z$, for $i=1,\ldots,r$ and $\gamma_i = \frac{p_i+1}{2}$ or $\frac{p_i+3}{4}$.

An isomorphism regarding \textit{Theorem \ref{crtquaternio}} for $O=\Hu$ can be obtained according to Bézout identity. Since $\text{GCD}(p_1, p_2,\ldots, p_r)=1$, there exist $\zeta_1,\ldots,\zeta_n\in\Hu$ such that for $\nu_i=q/p_i$, we have
        \begin{equation*}
            \nu_1 \zeta_1 + \nu_2 \zeta_2 + \ldots + \nu_k \zeta_k = 1.
        \end{equation*}

    Therefore, we can consider the isomorphism
        \begin{align*}
        \phi^{-1}:& \Hu/p_1\Hu\times\ldots\times\Hu/p_r\Hu \rightarrow \Hu/q\Hu \\
        \phi^{-1}(\hat{a}_1, \ldots,\hat{a}_r) &= (\nu_1 \zeta_1 \hat{a}_1+ \ldots + \nu_k \zeta_k \hat{a}_r)\hspace{-.2cm}\mod{q\Hu}.
        \end{align*}

    Replacing each $\hat{a}_i = \Psi^{-1}_i(a_i^{(1)}, a_i^{(2)})$, we arrive at the ring isomorphism
    {\small \begin{align}
        \varphi^{-1}(a_1^{(1)},a_1^{(2)},&\ldots,a_k^{(1)},a_k^{(2)}) = \nonumber\\ 
        & = \left[\sum_{i=1}^r \nu_i\zeta_i\gamma_i (\overline{\pi}_i a_i^{(1)} + \pi_i a_i^{(2)})\right]\hspace{-.2cm}\mod q\Hu.
\label{isomorfismohurwitz}
\end{align}}

Note that $\varphi$ is the standard modulo map,
\begin{align*}
   \varphi: \Hu/q\Hu &\rightarrow \Hu/\Hu\pi_1\times\ldots
   \times\Hu/\Hu\overline{\pi}_r\\
   a &\mapsto (a\mod\Hu\pi_1, \ldots, a\mod\Hu\overline{\pi}_r).
\end{align*}

The ring isomorphism defined in (\ref{isomorfismohurwitz}) can be used in Construction $\pi_A$ over Hurwitz integers described as follows:

\begin{Def}[Construction $\pi_A$ over Hurwitz integers]
	\label{construcao_pia_H}
	Let $\Hu$ be the maximal order of Hurwitz integers, let $p_1,\ldots,p_r$ be distinct odd rational primes and let $q=\prod_{j=1}^kp_j$. %, such that $\Hu_{p_j} = \Hu/p_j\Hu$ for $j=1,\ldots,r$, and $p_1\Hu,\ldots,p_r\Hu$ be distinct prime ideals of $\Hu$.
 Let $m_j$ and $n$ be integers such that $m_j^{(i)}\leq n, i=1,2$, and let $G_j^{(1)}:n\times m_j^{(1)}$ be a generator matrix of a linear code in $(\Hu/\Hu\pi_j)^n$ and $G_j^{(2)}:n\times m_j^{(2)}$ a generator matrix of a linear code in $(\Hu/\Hu\overline{\pi}_j)^n$ for $j\in\{1,\ldots,r\}$. Construction $\pi_A$ over Hurwitz integers consists of the following steps:
	\begin{enumerate}
		\item Define the discrete codebooks $\cod^{(1)}_j=\{G_j^{(1)}\cdot u: u\in (\Hu/\Hu\pi_j)^{m_j^{(1)}}\}$ and $\cod^{(2)}_j=\{G_j^{(2)}\cdot u: u\in (\Hu/\Hu\overline{\pi}_j)^{m_j^{(2)}}\}$ for $j\in\{1,\ldots,r\}$.
		\item Construct $\cod = \varphi^{-1}(\cod^{(1)}_1, \cod^{(2)}_1,\ldots,\cod^{(1)}_r,\cod^{(2)}_r)\subset (\Hu/q\Hu)^n$, where $\varphi : (\Hu/q\Hu )^n \rightarrow (\Hu/\Hu\pi_1)^n \times (\Hu/\Hu\overline{\pi}_1)^n \times \ldots \times (\Hu/\Hu\pi_k)^n \times (\Hu/\Hu\overline{\pi}_r)^n$ is a ring isomorphism.
		\item ``Expand'' $\cod$ to the entire space $\Hu^n$ to form the lattice $\displaystyle\Lambda_{\pi_A}^{\Hu}(\cod)=\cod+ q\Hu^n$.
	\end{enumerate}
\end{Def}

It is clear that Construction $\pi_A$ over Hurwitz integers always produces a lattice in $\Hu^n\subset\R^{4n}$.

\begin{Ex}
\label{exemplo_pia_quaternio}
Let \( q = 3 \) and \( n = 1 \). Consider the following ring isomorphism,
\[
\varphi^{-1}:\Hu/\Hu(1+\ii+\jj)\times\Hu/\Hu(1-\ii-\jj)\rightarrow \Hu/3\Hu.
\]
From equation~(\ref{eqcrt2}), since \( \pi = 1+\ii+\jj \) and \( \text{Re}(\pi) = 1 \), we compute \(\gamma = \frac{3+1}{2} = 2 \). Then the inverse isomorphism is given by,
\[
\varphi^{-1}(a^{(1)}, a^{(2)}) = 2\left[(1-\ii-\jj) a^{(1)} + (1+\ii+\jj) a^{(2)}\right] \mod{3\Hu},
\]
\noindent where \( a^{(1)} \in \Hu/\Hu\pi \) and \( a^{(2)} \in \Hu/\Hu\overline{\pi} \).

We can now apply \( \varphi^{-1} \) to construct a Construction \( \pi_A \) lattice over the Hurwitz integers. Consider the following linear codes
\begin{align*}
\cod_1 &= \left\{\frac{-1-\ii-\jj+\kk}{2}\cdot u \mod \Hu(1+\ii+\jj) \;:\; u\in \F_9\right\}\\
 & =\left\{0,\kk,\frac{-1-\ii-\jj+\kk}{2}\right\}\subset \Hu/\Hu\pi,\\
\cod_2 &= \left\{\frac{-1-\ii+\jj-\kk}{2}\cdot u \mod \Hu(1-\ii-\jj) \;:\; u \in \F_9\right\} \\
 & = \left\{0, j, \frac{-1-\ii+\jj-\kk}{2}\right\}\subset \Hu/\Hu\overline{\pi}.
\end{align*}

We have \(\cod = \varphi^{-1}(\cod_1\times \cod_2) = \{0, -1+\ii, 1-\ii, -1+\jj+\kk, 1-\jj-\kk, -\ii+\jj+\kk, \ii-\jj-\kk, \frac{1+\ii+\jj+\kk}{2}, \frac{1-\ii-\jj-\kk}{2}\}\subset \Hu/3\Hu \). Then the associated Construction \( \pi_A \) lattice is given by:
\[
\Lambda_{\pi_A}^{\Hu}(\cod) = \cod + 3\Hu^2 \subset \R^4,
\]
\noindent which represents the lattice generated by the code \( \cod \) via Construction \( \pi_A \).
Note that in this example, \( |\cod| = 9 = 3^2 \), and \( |\cod_1| = |\cod_2| = 3 \). 

A geometric visualization of the codes $\cod_1, \cod_2$ and $\cod$, is presented in Figure~\ref{examplecodes} via a projection onto the imaginary subspace of the quaternions, i.e., onto $\R^3$ by omitting the real part of each quaternion.

\begin{figure*}[htpb]
    \centering
    \subfloat[]{\includegraphics[width=0.33\linewidth]{cod1.pdf}}
    \subfloat[]{\includegraphics[width=0.33\linewidth]{cod2.pdf}}
    \subfloat[]{\includegraphics[width=0.33\linewidth]{cod0.pdf}}
    \caption{Projections of the codes of Example \ref{exemplo_pia_quaternio} onto $\R^3$ by omitting the real part. Gray points represent the projection of all 81 elements of $\Hu/3\Hu$. In (a), the code $\cod_1\subset\Hu/\Hu\pi$ is shown in blue; in (b), the code $\cod_2\subset\Hu/\Hu\overline{\pi}$ is shown in red, and, in (c) the code $\cod=\varphi^{-1}(\cod_1\times\cod_2)\subset\Hu/3\Hu$, obtained via Construction $\pi_A$, is represented by black points. The cube serves as a spatial reference in $\R^3$.}
    \label{examplecodes}
\end{figure*}
\end{Ex}

\begin{Obs}
A construction that exploits the natural order of Lipschitz was proposed in \cite{huang2018layered}. In that work, space-time index codes were designed for a $2\times 1$ Multiple-Input Single-Output (MISO) channel, based on the Alamout code. There it was considered the principal ideal domain $\Z[\ii]$ of Gaussian integers, elements $\phi_1,\ldots, \phi_r$ chosen as pairwise coprime, with $q=\prod_{i=1}^r\phi_i$ and $\text{Nrm}(\phi_i)=q_i$ for $i=1,\ldots, r$ where $q_i$ is not necessarily a prime number. Through CRT, it was established an isomorphism,
\[\vartheta:\Z[\ii]/q\Z[\ii] \cong \Z[\ii]/\phi_1\Z[\ii]\times\cdots\times \Z[\ii]/\phi_r\Z[\ii],\]
which enables representing, for each $i=1,\ldots, r$, a codeword $w_i\in(\Z[\ii]/\phi_i\Z[\ii])^2$ by two components, $w_i=(w_{i,0}, w_{i,1})$ with $w_{i,l}\in \Z[\ii]/\phi_i\Z[\ii], l=0,1$. The encoder maps the components $w_{1,l}, \ldots, w_{r,l}$ into the signal for the layer $l\in\{0,1\}$ via
\[
x_l = \vartheta(w_{1,l},\ldots, w_{r,l})\in\Z[\ii]/q\Z[\ii],
\]
with the overall codebook being a subset of $\cod_{\Z[\ii]}$ given by
\[
\cod_{\Z[\ii]} = \left\{\begin{pmatrix}
    x_0 & -\overline{x_1}\\
    x_1 & \overline{x_0}
\end{pmatrix}; x_0, x_1\in\Z[\ii]/q\Z[\ii]\right\}.
\]

In comparison to this approach, the construction described in \textit{Definition \ref{construcao_pia_H}} does not rely on a layered structure, offering a different approach. This allows the construction of codes using any odd rational prime $p$, enlarging the possibilities compared to the previous framework restricted to Gaussian primes.
\end{Obs}

\begin{Obs}
An interesting direction to be considered is the extension of our approach to the maximal order over the octonions $\mathbb{O}$, known as the \textit{Cayley integers} or \textit{Octavian integers} \cite{conway2003quaternions}. The Octavian integers, denoted by $\mathcal{O}$, are closely related to the notable $E_8$ lattice and form a non-commutative, non-associative, but yet \textit{alternative}\footnote{An alternative algebra is a non-associative algebra that must satisfies: $x(xy)=(xx)y$ and $(yx)x = y(xx)$ for all $x$ and $y$ in the algebra \cite[p.72]{conway2003quaternions}.} algebra. This alternative structure preserves enough algebraic properties to enable meaningful arithmetic.

In $\mathcal{O}$, a division algorithm and an ideal theory can still be developed, despite the non-associative nature of the algebra. As shown in \cite[pp. 109-110]{conway2003quaternions}, every left or right ideal in $\mathcal{O}$ is principal, and in fact, all ideals are two-sided. Moreover, any two-sided ideal in $\mathcal{O}$ is of the form $n\mathcal{O}$, where $n$ is a rational integer.

These properties highlight the strong arithmetic structure present in $\mathcal{O}$, enabled by the high symmetry inherited from the $E_8$ lattice. Moreover, a version of the CRT can be formulated in this setting, though it mirrors the classical CRT over $\Z$, as remarked in \cite{natarajan2015lattice}.

A significant challenge in extending our construction to this setting lies in characterizing conditions under which a CRT-type factorization, similar to Theorem~\ref{crt2}, can be applied to quotient rings of the form $\mathcal{O}/p\mathcal{O}$, where $p=\text{Nrm}(\theta)$ for some Octavian prime $\theta\in\mathcal{O}$. Understanding the arithmetic and algebraic structure of such quotients will be a key step to extends our framework to octonions settings.
\end{Obs}

\section{Multilevel Decoding Process}

In this section, we will describe a decoder that takes advantage of the layered structure of Construction $\pi_A$ lattices. This decoder was adapted from \cite{huang2017construction} where it is called serial modulo decoder (SMD). It is motivated by a decoding algorithm of Construction $D$ lattices as described by \cite{forney2000sphere} and \cite{matsumine2018construction}.

Let $y\in\R^{4n}$ be the received point, 
\[y=x+z,\]
where $x\in\Lambda_{\pi_A}^{\Hu}(\cod)$, and $z\in\R^{4n}$ represents the additive white Gaussian noise with variance $\sigma^2$. Since $x$ belongs to Construction $\pi_A$ lattice, it can be decomposed as	
\begin{align}
x&=\left[\sum_{i=1}^r \nu_i\zeta_i\gamma_i (\overline{\pi}_i a_i^{(1)} + \pi_i a_i^{(2)})\hspace{-.2cm}\mod q\Hu\right] + q\tilde{\lambda},
\label{dec1}
\end{align}

\noindent where $q=\prod_{l=1}^r p_l$ and $\tilde{\lambda}\in\Hu^n$.

Consider $\mathcal{B}=\{v_1,\ldots,v_{4n}\}$ a basis of $\Hu^n$ and the fundamental region $\Partopo(\mathcal{B})$ of $q\Hu^n$ in $\R^{4n}$, $\Partopo(\mathcal{B})=\{\sum_{i=1}^{4n} a_i v_i, a_i\in[-q/2,q/2)\},$ then reducing $y$ to $\Partopo(\mathcal{B})$, denoted by $y\mod q\Hu$, we get
    \[ y\mod q\Hu = y - q\hat{\lambda} =: y_1^{(1)},\]

    \noindent where $\hat{\lambda}=\lfloor q^{-1}\cdot y \rceil$.

    Then, 
    {\small \begin{align*}
    y_1^{(1)} = \left[\sum_{i=1}^r \nu_i\zeta_i\gamma_i (\overline{\pi}_i a_i^{(1)} + \pi_i a_i^{(2)})\hspace{-.2cm}\mod q\Hu\right]+ q \lambda + z,
     \end{align*}}

    \noindent where $\lambda:=\tilde{\lambda}-\hat{\lambda}\in\Hu^n$.

Operate with $\cod_1^{(1)}$ applying $\mod\pi_1\Hu$ to the received sequence $y_1^{(1)}$,
	{\small \begin{align*}
	y_1'^{(1)}:= & y_1^{(1)}\mod \pi_1\Hu = \nu_1\zeta_1 \gamma_1\overline{\pi}_1 a_1^{(1)}\mod \pi_1\Hu + z_1\\
	=& a_1^{(1)}\mod \pi_1\Hu + z_1^{(1)},
	\end{align*}}
	
	\noindent where $z_1^{(1)}:=z\mod \pi_1\Hu$.
 
Apply a decoder to $y_1'^{(1)}$ to obtain the codeword $a_1^{(1)}\in\cod_1^{(1)}$. Once we have $a_1^{(1)}$, perform a re-encoding process by
    \begin{equation*}
        y_1^{(2)}:=y_1^{(1)}-\nu_1\zeta_1\gamma_1\overline{\pi}_1 a_1^{(1)}.
        \label{reencoding1}
    \end{equation*}

 Repeat the same steps, now operating with $\cod_1^{(2)}$, applying $\mod\overline{\pi}_1\Hu$ to $y_1^{(2)}$ to get $a_1^{(2)}$ and then re-encoding again to get
    \begin{equation*}
        y_2^{(1)}:=y_1^{(2)}-\nu_1\zeta_1\gamma_1\pi_1 a_1^{(2)}.
    \end{equation*}

    Repeat these steps for all levels $L=2,\ldots,r$.
   
	In the last step $r+1$, we have
    \begin{equation*}
    y_{r+1}= z+q\lambda \Rightarrow z=y_{r+1}\mod q\Hu, 
    \end{equation*}
	
	%Then we obtain $\lambda=y_{r+1}\mod q\Hu$ 
 Then we can recover $x$, the lattice point closest to the received $y$,
    \begin{equation*}
        x=y-z.
    \end{equation*}

Algorithm \ref{algo2} is the description of this process. We denote as $\text{Dec}_j, j=1,\ldots,r$, any decoder that identifies the closest point $x$ in the Construction A lattice from $\cod_i^{(j)}$. %Additionally, $\mod^*$ denotes the modulus reduction, ensuring that the residues are within the range of $[-q/2, q/2]$.

\begin{algorithm}
\caption{Construction $\pi_A$ lattices from Hurwitz integers decoder.}
\label{algo2}
\begin{algorithmic}
	\REQUIRE received message $y$.
%	$q=p_1\cdot \ldots\cdot p_r$;\\
%	\For{$i=1$ \KwTo$r$}{
%        $p_i=\pi_i\cdot\overline{\pi_i}$\\
%        \If{$Re(\pi_i)=1$}{$\gamma_i=(p_i+1)/2$}
%        \If{$Re(\pi_i)=2$}{$\gamma_i=(3p_i+1)/4$}
%		$\nu_i=q/p_i$;\\
%        \ForEach{$\nu_i$}{
%			$\nu_i\zeta_i\equiv1\mod p_i$;\\
%			\Return{$\zeta_i$}	
%		}
%	}
	\STATE $y_1=y\mod q\Hu$;
	\STATE $y_1^{(1)}=y_1$;
	\FOR{$j = 1$ to $r$}
	\STATE	$y_j'^{(1)}=y_j^{(1)}\mod \pi_j\Hu$;
	\STATE	$\hat{c}_j^{(1)}=\textrm{Dec}_j(y_j'^{(1)})$;
	\STATE	$y_{j}^{(2)}=y_j^{(1)}-\gamma_j\nu_j\zeta_j\overline{\pi}_j\hat{c}_j^{(1)}$;
	\STATE	$y_j'^{(2)}=y_j^{(2)}\mod \overline{\pi}_j\Hu$;
	\STATE	$\hat{c}_j^{(2)}=\textrm{Dec}_j(y_j'^{(2)})$;
	\STATE	$y_{j+1}^{(1)}=y_j^{(2)}-\gamma_j\nu_j\zeta_j\pi_j\hat{c}_j^{(2)}$;
    \ENDFOR
	%$\hat{x}=(x_1 m_1 \hat{c}_1 + \ldots+ x_k m_k \hat{c}_r)\mod q$;\\
	\STATE $z=y_{r+1}\mod q\Hu$;
	%	$z=\left\lfloor\frac{y_{r+1}-z}{q}\right\rfloor$;\\
	\STATE$x= y - z$
 \RETURN estimated lattice point $x\in\Lambda_{\pi_A}^{\Hu}(\cod)$.
 \end{algorithmic}
\end{algorithm}

\subsection{Decoding complexity}

Using Construction $\pi_A$ over Hurwitz integers, we can enhance the decoding complexity for codes of ``large'' sizes. It is noteworthy that, related to Construction $\pi_A$ lattices over $\mathbb{Z}$, $\mathbb{Z}[\mathrm{i}]$, $\mathbb{Z}[\omega]$, \cite{huang2017construction}, the Construction $\pi_A$ over $\mathscr{H}$ for $\mathcal{C} = \varphi^{-1}(\mathcal{C}_1^{(1)} \times \ldots \times \mathcal{C}_r^{(2)})$ has decoding complexity influenced by the greatest cardinality of codes $\mathcal{C}_1^{(1)}, \mathcal{C}_1^{(2)}, \ldots, \mathcal{C}_r^{(1)}, \mathcal{C}_r^{(2)}$, denoted as $M = \max\{|\mathcal{C}_i^{(1)}|, |\mathcal{C}_i^{(2)}|, i = 1, \ldots, r\}$, roughly $O(M\log_2 M)$.

Table \ref{table_time} presents a comparison of the computational time required to decode linear codes generated through Construction $\pi_A$ over $(\mathbb{Z}/q\mathbb{Z})^4$, $(\mathbb{Z}[\mathrm{i}]/q\mathbb{Z}[\mathrm{i}])^2$, $(\mathbb{Z}[\omega]/q\mathbb{Z}[\omega])^2$, and $\mathscr{H}/q\mathscr{H}$. The decoding was performed using the serial modulo decoding algorithm. For a fair comparison, all codes considered have the same size, $q^4$ elements in the respective quotient ring, and we used the canonical basis vectors as generators in each case. A notable improvement in computational efficiency is observed in some cases when the decoding over $\mathscr{H}$ is lower compared to the other rings. However, it is worth mentioning that for $q=39$, where the maximum code size in both $\Z[\ii]/q\Z[\ii]$ and $\Hu/q\Hu$ is $13^2$, the decoding over $\Z[\ii]$ requires only $0.54506$ seconds, while decoding over $\Hu$ takes $1.22351$ seconds. All simulations were performed using an Intel\textregistered \ Xeon\textregistered \ Gold 6154 CPU.

\begin{table}[htpb]
		\centering
		\caption{{\small Decoding time comparison between Construction $\pi_A$ lattices from $\mathbb{Z}$, $\mathbb{Z}[\mathrm{i}]$, $\mathbb{Z}[\omega]$, and $\mathscr{H}$.}}
  \label{table_time}
\begin{tabular}{c|c|cccc}
    \hline
          &  &  \multicolumn{4}{c}{Time (seconds)}  \\
    \cline{3 - 6}
    \multirow{-2}{*}{q}   & \multirow{-2}{*}{Code Size} & $\mathbb{Z}$ & $\mathbb{Z}[\mathrm{i}]$ & $\mathbb{Z}[\omega]$ & $\mathscr{H}$\\
			\hline \hline
			15 & $50625$   & $3.50723$ & $0.080691$ &    --   & $\textbf{0.060011}$ \\
			21 & $194481$  & $62.5559$  &   --    & $0.53549$ & $\textbf{0.234559}$ \\
			33 & $1185921$ & $2600.66$  &   --    &   --    & $\textbf{0.681975}$ \\
			35& $1500625$  & $66.3755$  & $100.837$ & $21.7788$ & $\textbf{0.663997}$ \\
            39& $2313441$  & $8718.13$  & $\textbf{0.54506}$ &  --     & $1.22351$  \\
            55& $9150625$  & $2328.67$  & $4014.38$ &    --   & $\textbf{0.684826}$ \\
%            69& $22667121$ & Running  &   --    &     --  & Running  \\
            77& $35153041$ & $2576.51$  &  --     & $15284.9$ & $\textbf{0.762427}$ \\
			\hline
		\end{tabular}
\end{table}

As remarked in \cite{huang2017construction}, for Construction A lattices the decoding complexity is dominated by $|\cod|$. In the context of a Construction $\pi_A$ multilevel decoder, maximum-likelihood decoding occurs independently at each level. Consequently, the decoding process is dominated by the largest cardinality of the codes involved in the encoding.

In the general case, from the isomorphism
\[(\ring/q\ring)^n\cong(\ring/p_1\ring)^n\times\cdots \times(\ring/p_r\ring)^n,\]

\noindent with $q=p_1\cdot \ldots \cdot
 p_r$, where $p_i$ are prime elements over $\ring=\Z, \Z[\ii], \Z[\ww]$, or $\Hu$, we have
\[\cod=\varphi^{-1}(\cod_1\times \cdots\times\cod_r)\subset(\ring/q\ring)^n,\]

\noindent where $\cod_i\subset(\ring/p_i\ring)^n$ represents a linear code of size $|\cod_i|$, and $|\cod|=\prod_{i=1}^r |\cod_i|$. Thus, $\cod$ is associated with a lattice code in $\R^n, \R^{2n}, \R^{2n}$, and $\R^{4n}$ for $\ring=\Z, \Z[\ii], \Z[\ww]$, and $\Hu$, respectively. For LDPC codes in each level, the decoding complexity of Construction $\pi_A$ is reported to be roughly  $O(\cod_{\text{max}}\log_2 \cod_{\text{max}})$, where $\cod_{\text{max}}=\max\{|\cod_i|\}$, \cite{huang2017construction}, \cite{davey1998}.

It's worth noting that using lattices constructed from Hurwitz integers also expands the design space. For instance, for a code $\cod$ with $|\cod|=81$, there is a Construction $\pi_A$ over the Hurwitz integers, but there is no  Construction $\pi_A$ over $\Z, \Z[\ii]$, or $\Z[\ww]$ due to the lack of a partition into distinct prime ideals for a code of this cardinality. Moreover, also for $p\equiv 11\mod 12$, a partition always exists for Hurwitz integers but codes with cardinality $p^4$ cannot be obtained by Construction $\pi_A$ over $\Z, \Z[\ii]$ or $\Z[\ww]$.

Additionally, considering codes $\cod=(\ring/q\ring)^{4n}$ obtained through Construction $\pi_A$ in dimension $4n$, the computational complexity of decoding the lattice codes over $\Z, \Z[\ii]$, and $\Z[\ww]$ do not exceed those obtained through $\Hu$ as detailed next.

%We will analyse the case where the code $\cod$ is the whole set $(\ring/q\ring)^n$.
For $\ring=\Z$ and $q=p_1\cdot p_2\cdot\ldots\cdot p_r$ we have, %the computational complexity is approximately $O(p_{\text{max}}^{4n}\log_2 p_{\text{max}}^{4n})$, where $p_{\text{max}} = \max\{ p_i\},$ $i=1,\cdots,r$.
\[
\left(\frac{\Z}{q\Z}\right)^{4n}\cong  \left(\frac{\Z}{p_1\Z}\right)^{4n}\times\left(\frac{\Z}{p_2\Z}\right)^{4n}\times\cdots\times \left(\frac{\Z}{p_r\Z}\right)^{4n},
\]

\noindent and the computational complexity is $O(p_{\Z}\log_2 p_{\Z})$, where $p_{\Z}=\max\{p_i^{4n}\}$, $i=1,\ldots,r$.

For the Gaussian integers, $\ring = \Z[\ii]$, we can rearrange $q = p_1\cdot\ldots\cdot p_s\cdot p_{s+1}\cdot\ldots\cdot p_r$, with $ p_i\equiv 1\mod 4, i=1,\ldots,s$ and $p_j\equiv 3\mod 4, j=s+1,\ldots,r$. Since $\Z[\ii]/p_i\Z[\ii]$ is factorable in non trivial rings only for $i=1,\ldots,s$, i.e, $(\Z[\ii]/p_i\Z[\ii])^{2n} = (\Z[\ii]/\pi_i\Z[\ii])^{2n}\times (\Z[\ii]/\overline{\pi}_i\Z[\ii])^{2n}$, we have 
{\small \begin{align*}
\left(\frac{\Z[\ii]}{q\Z[\ii]}\right)^{2n}\cong & \left(\frac{\Z[\ii]}{p_1\Z[\ii]}\right)^{2n}\times\ldots\times\left(\frac{\Z[\ii]}{p_s\Z[\ii]}\right)^{2n}\times\\
& \times\left(\frac{\Z[\ii]}{p_{s+1}\Z[\ii]}\right)^{2n}\times\ldots\times \left(\frac{\Z[\ii]}{p_r\Z[\ii]}\right)^{2n}\\
 \cong & \left(\frac{\Z[\ii]}{\pi_1\Z[\ii]}\right)^{2n}\hspace{-.2cm}\times\left(\frac{\Z[\ii]}{\overline{\pi}_1\Z[\ii]}\right)^{2n}\hspace{-.2cm}\times\ldots\times \left(\frac{\Z[\ii]}{\pi_r\Z[\ii]}\right)^{2n}\hspace{-.2cm}\times\\
& \left(\frac{\Z[\ii]}{\overline{\pi}_r\Z[\ii]}\right)^{2n}\times\left(\frac{\Z[\ii]}{p_{s+1}\Z[\ii]}\right)^{2n}\times\ldots\times \left(\frac{\Z[\ii]}{p_r\Z[\ii]}\right)^{2n}\hspace{-.2cm},
\end{align*}}

\noindent the computational complexity is $O(p_{G}\log_2 p_{G})$, where $p_{G} = \max\{p_i^{2n}, p_j^{4n},i=1,\ldots,s, \ j=s+1,\ldots,r\}$. We can proceed in an analogous way in the case of Eisenstein integers, by rearranging $q = p_1\cdot\ldots\cdots p_{\tilde{s}}\cdot p_{\tilde{s}+1}\cdot\ldots\cdot p_r$, with $ p_i\equiv 1\mod 3, i=1,\ldots,\tilde{s}$ and $p_j\equiv 2\mod 3, j=\tilde{s}+1,\ldots,r$. Again, we have that $(\Z[\ww]/p_i\Z[\ww])^{2n}$ is only factorable on non trivial rings for $i=1,\ldots,\tilde{s}$, and we can assure that the computational complexity is $O(p_{E}\log_2 p_{E})$, where $p_{E} = \max\{p_i^{2n}, p_j^{4n}, i=1,\ldots,\tilde{s}, \ j=\tilde{s}+1,\ldots,r\}$.

Meanwhile, in $\Hu$, for $q=p_1\cdot\ldots\cdot p_r$, where $p_i=N(\pi_i)$ and $\pi_i$ is a irreducible element in $\Hu$, we have
{\small\begin{align*}
\left(\frac{\Hu}{q\Hu}\right)^{n}& \cong \left(\frac{\Hu}{\Hu p_1}\right)^{n}\times\ldots\times \left(\frac{\Hu}{\Hu p_r}\right)^{n}\\
 & \cong \left(\frac{\Hu}{\Hu\pi_1}\right)^{n}\hspace{-.2cm}\times\left(\frac{\Hu}{\Hu\overline{\pi}_1}\right)^{n}\hspace{-.2cm}\times\ldots\times \left(\frac{\Hu}{\Hu\pi_r}\right)^{n}\hspace{-.2cm}\times\left(\frac{\Hu}{\Hu\overline{\pi}_r}\right)^{n},
\end{align*}}

\noindent with computational complexity $O(p_{\Hu} \log_2 p_{\Hu})$, where $p_{\Hu} =\max\{p_i^{2n}\}$, $i=1,\ldots,r$.

Comparing the four scenarios we have,
\begin{enumerate}
    \item $O(p_{\Hu}\log_2 p_{\Hu}) < O(p_{\Z}\log_2 p_{\Z})$;
    \item $O(p_{\Hu}\log_2 p_{\Hu}) \leq O(p_{G}\log_2 p_{G})$ and equality holds only if we have for $p=\max\{{p_i}\}$, $1\leq i\leq r$, $p\equiv 1\mod 4$;
    \item $O(p_{\Hu}\log_2 p_{\Hu}) \leq O(p_{E}\log_2 p_{E})$ and equality holds only if we have for $p=\max\{{p_i}\}$, $1\leq i\leq r$, $p\equiv 1\mod 3$;
\end{enumerate}

Table \ref{table_complexity} and Figure \ref{decoding_complexity} compare the computational complexity of decoding in Construction $\pi_A$ over $\Z[\ii]$, $\Z[\ww]$, and $\Hu$, in real dimension $4$ (i.e. $n=1$) and for constellations with cardinality $q^4$.

\begin{figure}[htpb]
\centering
\includegraphics[scale=.248]{complexity_graphic.pdf}
\caption{Comparison of decoding complexity for constellations with cardinality $q^4$ in $\R^4$ obtained through Construction $\pi_A$ over $\Z[\ii]$, $\Z[\ww]$, and $\Hu$. Here, $\cod_{\text{max}}$ denotes the maximum cardinality among the layers in the encoding.}
\label{decoding_complexity}
\end{figure}

\begin{table}[htpb]
    \renewcommand{\arraystretch}{1.2}
    \caption{Factorization of constellations in $\R^4$ over $\Z[\ii], \Z[\ww]$ and $\Hu$.}
    \label{table_complexity}
    \footnotesize
    \centering
    \begin{tabular}{>{\centering}m{0.04\columnwidth}>{\centering}m{0.07\columnwidth}>{\centering}m{0.5\columnwidth}>{\centering}m{0.08\columnwidth}}
    \hline
    $q$ & \multicolumn{2}{c}{Factorization} & $\cod_{\text{max}}$ \\
    \hline
    \hline
        3   & $\Hu$ & $(1+\ii+\jj)\cdot(1-\ii-\jj)$ & $3^2$ \\
    \hline
          & $\Z[\ii]$ &   \\
    %\cline{3 - 4}
    \multirow{-2}{*}{5}   & $\Hu$ & \multirow{-2}{*}{$(1+2\ii)\cdot(1-2\ii)$} & \multirow{-2}{*}{$5^2$}\\
    \hline
          & $\Z[\ww]$ & $(1+3\ww)\cdot (-2-3\ww)$ & \\
    %\cline{3 - 4}
    \multirow{-2}{*}{7}   & $\Hu$ & $(1+\ii+\jj+2\kk)\cdot(1-\ii-\jj-2\kk)$ & \multirow{-2}{*}{$7^2$}\\
    \hline
    11   & $\Hu$ & $(1+\ii+3\jj)\cdot(1-\ii-3\jj)$ & $11^2$\\
    \hline
       & $\Z[\ii]$ & $(2+3\ii)\cdot (2-3\ii)$ & \\
       & $\Z[\ww]$ & $(1+4\ww)\cdot (-3-4\ww)$ & \\
    %\cline{3 - 4}
    \multirow{-3}{*}{13}   & $\Hu$ & $(1+2\ii+2\jj+2\kk)\cdot(1-2\ii-2\jj-2\kk)$ & \multirow{-3}{*}{$13^2$}\\
    \hline
%      & $\Z$ & $3\cdot 5$ & $O(5^4\log_2 5^4)$ \\
    %\cline{3 - 4}
        & $\Z[\ii]$ & $3\cdot (1+2\ii)\cdot (1-2\ii)$ & $3^4$\\
    %\cline{3 - 4}
    \multirow{-2}{*}{15}   & $\Hu$ & $(1+\ii+\jj)(1-\ii-\jj)(1+2\ii)(1-2\ii)$ & $5^2$\\
    \hline
         & $\Z[\ii]$ &  & \\
    %\cline{3 - 4}
    \multirow{-2}{*}{17}   & $\Hu$ & \multirow{-2}{*}{$(1+4\ii)\cdot (1-4\ii)$} & \multirow{-2}{*}{$17^2$}\\
    \hline
          & $\Z[\ww]$ & $(2+5\ww)\cdot (-3-5\ww)$ & \\
    %\cline{3 - 4}
    \multirow{-2}{*}{19}   & $\Hu$ & $(1+\ii+\jj+4\kk)\cdot(1-\ii-\jj-4\kk)$ & \multirow{-2}{*}{$19^2$}\\
    \hline
%          & $\Z$ & $3\cdot 7$ & $O(7^4\log_2 7^4)$ \\
    %\cline{3 - 4}
    & & $(1+\ii+\jj) (1-\ii-\jj) (1+\ii+\jj+2\kk)$ & \\
    \multirow{-2}{*}{21}   & \multirow{-2}{*}{$\Hu$} & $(1-\ii-\jj-2\kk)$  & \multirow{-2}{*}{$21^2$}\\
    \hline
        23   & $\Hu$ & $(1+2\ii+2\jj+3\kk)\cdot(1-2\ii-2\jj-3\kk)$ & $23^2$\\
    \hline
          & $\Z[\ii]$ & $(2+5\ii)\cdot(2-5\ii)$ & \\
    %\cline{3 - 4}
    \multirow{-2}{*}{29}   & $\Hu$ & $(2+3\ii+4\jj)\cdot (2-3\ii-4\jj)$ & \multirow{-2}{*}{$29^2$}\\
    \hline
         & $\Z[\ww]$ & $(1+6\ww)\cdot (-5-6\ww)$ & \\
    %\cline{3 - 4}
    \multirow{-2}{*}{31}   & $\Hu$ & $(1+\ii+2\jj+5\kk)\cdot(1-\ii-2\jj-5\kk)$ & \multirow{-2}{*}{$31^2$}\\
    \hline
%          & $\Z$ & $3\cdot 11$ & $O(11^4\log_2 11^4)$ \\
    %\cline{3 - 4}
       & & $(1+\ii+\jj)(1-\ii-\jj)(1+\ii+3\jj)$ & \\
    \multirow{-2}{*}{33}   & \multirow{-2}{*}{$\Hu$} & $(1-\ii-3\jj)$ & \multirow{-2}{*}{$11^2$}\\
    \hline
%       & $\Z$ & $5\cdot 7$ &  \\
    %\cline{3 - 4}
       & $\Z[\ii]$ & $7\cdot (1+2\ii)\cdot (1-2\ii)$ & $7^4$\\
       & $\Z[\ww]$ & $5\cdot (1+3\ww)\cdot (-2-3\ww)$ & $5^4$\\
       &  & $(1+2\ii)(1-2\ii)(1+\ii+\jj+2\kk)$ & \\
    \multirow{-4}{*}{35}   & \multirow{-2}{*}{$\Hu$} & $(1-\ii-\jj-2\kk)$ & \multirow{-2}{*}{$7^2$}\\
    \hline
       & $\Z[\ii]$ & $(1+6\ii)\cdot (1-6\ii)$ & \\
       & $\Z[\ww]$ & $(3+7\ww)\cdot (-4-7\ww)$ & \\
    %\cline{3 - 4}
    \multirow{-3}{*}{37}   & $\Hu$ & $(1+2\ii+4\jj+4\kk)\cdot(1-2\ii-4\jj-4\kk)$ & \multirow{-3}{*}{$37^2$}\\
    \hline
%      & $\Z$ & $3\cdot 13$ & $O(13^4\log_2 13^4)$ \\
    %\cline{3 - 4}
        & $\Z[\ii]$ & $3\cdot (2+3\ii)\cdot (2-3\ii)$ & \\
    %\cline{3 - 4}
      &  & $(1+\ii+\jj)(1-\ii-\jj)(1+2\ii+2\jj+2\kk)$ & \\
    \multirow{-3}{*}{39} & \multirow{-2}{*}{$\Hu$} & $(1-2\ii-2\jj-2\kk)$ & \multirow{-3}{*}{$13^2$}\\
    \hline
    \end{tabular}
\end{table}

\begin{Obs}
It is worth noting that, when considering general codes $\cod\subset(\Hu/q\Hu)^n\subset\R^{4n}$, it is possible to achieve lower decoding complexity for other constellations in $\R^{4n}$ with the same cardinality by using Gaussian or Eisenstein integers rather than Hurwitz integers.

For instance, in $\R^8$, if we consider
\begin{equation*}
\left(\frac{\Hu}{\Hu(1+2\ii)}\right)^2\times \left(\frac{\Hu}{\Hu(1-2\ii)}\right)^2 \rightarrow \left(\frac{\Hu}{5\Hu}\right)^2,
\label{H5H}
\end{equation*}

\begin{equation*}
    \left(\frac{\Z[\ii]}{\Z[\ii](1+2\ii)}\right)^4\times \left(\frac{\Z[\ii]}{\Z[\ii](1-2\ii)}\right)^4 \rightarrow \left(\frac{\Z[\ii]}{5\Z[\ii]}\right)^4,
    \label{G5G}
\end{equation*}

\noindent and codes $\cod_1\in (\Hu/\Hu(1+2\ii))^2$ and $\cod_2\in (\Hu/\Hu(1-2\ii))^2$ with $\text{rank}(\cod_1)=1$ and $\text{rank}(\cod_2)=2$, respectively, then using Construction $\pi_A$, we have a code $\cod\subset(\Hu/5\Hu)^2$ with cardinality $|\cod|=5^6$, and the decoding complexity is of order $5^4\log_2 5^4$. Conversely, a code $\cod\subset(\Z[\ii]/5\Z[\ii])^4$ with $|\cod|=5^6$ can be obtained by Construction $\pi_A$ over the Gaussian integers from codes with rank $3$ in each layer and therefore the decoding complexity is of order $5^3\log_2 5^3$.
\end{Obs}

Constellations of the Hurwitz integers have been considered in the context of four-dimensional coherent optical systems \cite{agrell2009, karlsson2011, zhou2016}. It has been shown that they can have good performance regarding shaping gain and different constrains for the shape such the hypercube, Voronoi region and spherical \cite{welti1974, stern2019hurwitz, frey2020,stern2022algorithms}. Particularly constellations of $2^b, b\in\N$ points are studied. %It seems to be interesting to investigate if the constellations $\Hu/q\Hu$ of $q^{4}$ points, $q=p_1\cdot\dots\cdot p_r, \ p_i$ prime, considered above, may enlarge this scenario also providing good shaping gain with lower computational complexity.\\

To evaluate the performance of existing constellations, Table~\ref{comparison} compares some parameters: the minimum distance $d_{\min}$, the average symbol energy ($E$), the normalized second moment (NSM), and the constellation figure of merit (CFM). In this comparison, it is considered Erik Agrell’s database of sphere packings and constellations \cite{agrell2019database}. For clarity, we briefly define each parameter below.

The average symbol energy $E$ quantifies the mean energy required to transmit a symbol and is defined as
\(
E = \frac{1}{M} \sum_{i=1}^M \|x_i\|^2,
\)
\noindent where $M$ denotes the number of points in the constellation, and $x_i\in\R^n$ is the vector representing the $i-$th constellation point. This quantity corresponds to the mean squared Euclidean norm of the constellation points.

The normalized second moment (NSM) measures the energy efficiency of a constellation relative to its minimum distance and it is given by,
\(
\text{NSM} = \frac{E}{d_{\min}^2}, 
\)
\noindent where $d_{\min}$ is the minimum Euclidean distance between any two distinct constellation points. Lower NSM values indicate tighter packings.

The constellation figure of merit (CFM) \cite{forney1989constellation} serves as a quality metric that balances minimum distance and average energy. It is defined as
\(
\text{CFM} = \frac{2 d_{\min}^2}{E}.
\)
\noindent and reflects how well-spread a constellation is with respect to energy usage. For an easy comparison, we report CFM in decibels (dB), computed as $10\log_{10}\text{CFM}$.

Constellations and sphere packings can be computed under many criteria depending on the communication scenario. When comparing constellations of the same dimension and cardinality, the one with lowest NSM is preferred, as it indicates higher energy efficiency. Alternatively, when comparing modulation formats under the same bandwidth, the format with the largest CFM offers better performance in terms of distance-to energy balance.

In Table \ref{comparison}, we adopt the notation used in \cite{agrell2019database} to refer to various classes of sphere packings in dimension 4. The entries w4\_* correspond to good packings of selected sizes originally proposed by Welti in \cite{welti1974}, obtained by selecting points from $D_4$ lattice that lie in certain chosen hyperspheres. The labels l4\_* refer to optimal subsets of the $D_4$ lattice (in terms of constellation design), constructed in unpublished work by Agrell. These subsets are not necessarily optimal sphere packings. Finally, the c4\_* entries represent conjectured optimal sphere packings drawn from G. Nebe and N. Sloane unpublished tables \cite{nebe2014database}. 

\begin{table}[htpb]
    \centering
    \caption{{\small Comparison of Average Symbol Energy, NSM, and CFM for Construction $\pi_A$ lattice codes via Hurwitz integers and existing approaches with low cardinality.}}
    \label{comparison}
  \begin{tabular}{p{1cm}  m{.6cm} m{1.1cm} m{1.1cm} m{1.1cm} m{1.2cm}}
    \hline
             & Code size             & $d_{\min}$ & $E$ & NSM & CFM (dB)  \\
          \hline \hline
tetra4     &                       & $1.732$  & $2.666$  & $0.889$ & $3.521$\\
l4\_9         &                       & $18$       & $192$      & $0.592$ & $5.282$\\
c4\_9         &                       & $1.608$  & $1.481$  & $\mathbf{0.572}$ & $\mathbf{5.433}$\\
$\Hu_{\pi_A}$& \multirow{-4}{*}{$9$}  & $1.732$  & $2.666$  & $0.889$ & $3.521$\\
        \hline
2prism4&                      & $1.175$  & $2$        & $1.447$ & $1.404$\\
l4\_25         &                      & $2$        & $3.84$     & $\mathbf{0.96}$    & $\mathbf{3.187}$\\
c4\_25         &                      & $2$        & $3.84$     & $\mathbf{0.96}$    & $\mathbf{3.187}$\\
$\Hu_{\pi_A}$& \multirow{-4}{*}{$25$} & $2.236$  & $4.8$      & $\mathbf{0.96}$    & $\mathbf{3.187}$\\
          \hline
% l4\_27         &                       & $54$       & $3000$     & $\mathbf{1.028}$ & $\mathbf{2.886}$\\
% c4\_27         &                       & $1.445$  & $2.150$  & $\mathbf{1.028}$ & $\mathbf{2.886}$\\
% $\Hu_{\pi_A}$& \multirow{-3}{*}{$27$}  & $1$        & $2.222$   & $2.222$  & $-0.457$\\
% \hline
2prism4&                       & $0.867$  & $2$        & $2.655$ & $-1.231$\\
w4\_49          &                       & $1.414$ & $2.938$  & $\mathbf{1.469}$ & $\mathbf{1.338}$\\
l4\_49          &                       & $2$       & $5.877$  & $\mathbf{1.469}$ & $\mathbf{1.338}$\\
$\Hu_{\pi_A}$& \multirow{-4}{*}{$49$}   & $2.645$ & $10.285$  & $\mathbf{1.469}$ & $\mathbf{1.338}$\\
\hline
l4\_81         &                        & $162$     & $49712$    & $\mathbf{1.894}$ &$\mathbf{0.235}$\\
$\Hu_{\pi_A}$& \multirow{-2}{*}{$81$}   & $1$       & $2.074$  & $2.074$ & $-0.157$\\
\hline
2prism4&                      & $0.563$ & $2$         & $6.299$ & $-4.982$\\
l4\_121         &                     & $242$      & $135936$    & $\mathbf{2.321}$ & $\mathbf{-0.646}$\\
$\Hu_{\pi_A}$& \multirow{-3}{*}{$121$}  & $3.316$  & $26.181$   & $2.380$ & $-0.755$\\
\hline
% l4\_125         &                       & $250$      & $147552$ & $\mathbf{2.360}$ & $\mathbf{-0.720}$\\
% $\Hu_{\pi_A}$& \multirow{-2}{*}{$125$}  & $1$        & $5.44$   & $5.44$    & $-4.345$\\
% \hline
w4\_169         &                       & $1$        & $5.396$  & $\mathbf{2.698}$ & $\mathbf{-1.300}$\\
$\Hu_{\pi_A}$& \multirow{-2}{*}{$169$}  & $3.605$   & $36.92$  & $2.840$ & $-1.523$\\
\hline
\end{tabular}
\end{table}

\section{Goodness for channel coding}

The main objective of this section is to show that Construction $\pi_A$ lattices from codes over $\Z, \Z[\ii], \Z[\ww]$ and $\Hu$ can be ``good for coding'' over the AWGN channel. In \cite{huang2017construction}, a proof with different approach, using arguments more closely to those of Forney \cite{forney2000sphere}, is presented for $\Z, \Z[\ii]$ and $\Z[\ww]$.

\subsection{Balanced set of codes over rings}

Let $\mathcal{R}$ be a finite ring, and $\ring^{\times}$ be its units. Denote by $(\ring^n)^{\times}$ the set of vectors in $\ring^n$ such that at least one coordinate is a unit.

In this section, we consider a linear code $\cod\subset\ring^n$ as a free\footnote{In the previous sections we have assumed the usual definition of linear codes over rings as a $\ring-$submodule. However, in this section, we require also that a linear code has a basis. These codes are called free linear codes \cite{dougherty2017algebraic}.} $\ring$-submodule of $\ring^n$. The rank $k$ of such a linear code $\cod$, $\text{rank}(\cod)=k$, is the number of elements of a basis of $\cod$ \cite{hartley1970modules}. We define a balanced set of codes as follows.

\begin{Def}[Balanced set of codes]
Consider a non-empty set of linear codes $\cod_{b}$ of the same cardinality. We say that $\cod_b$ is balanced in $\ring^n$ if every $x\in(\ring^n)^{\times}$ belongs to the same number, $L$, of codes of $\cod_b$.
\end{Def}

As an example, let $\ring=\Z_p$, $p$ prime. Since a nonvanishing element of $\Z_p$ is a unit, $(\Z_p^n)^{\times}=\Z_p^n-\{\mathbf{0}\}$. Therefore, 
\[
|(\Z_p^n)^{\times}|=p^n-1,
\]

Clearly, the set $\cod_{b_k}$ of all linear codes with rank $k$ over $\Z_p^n$ is a balanced set with
\[
L\cdot (p^{n}-1) = |\mathcal{C}_{b_k}| \cdot (p^k-1),
\]

\noindent where $|\cod_{b_k}|$ is the number of codes in $\cod_{b_k}$, \cite{loeliger1997averaging}. In fact, $|\cod_{b_k}|$ is the so called Gaussian binomial coefficient
\[
|\cod_{b_k}|=\binom{n}{k}_p = \dfrac{(p^n-1)(p^n-p)\dots(p^n-p^{k-1})}{(p^k-1)(p^k-p)\dots(p^k-p^{k-1})}.
\]

For a general finite ring $\ring$ with $q$ elements, we have that the set $\cod_{b_k}$ of all linear codes with rank $k$ is balanced with $|\cod|= M$ for $\cod\in\cod_{b_k}$ and
\begin{equation}
    L\cdot |(\ring^n)^{\times}| \leq |\cod_{b_k}| \cdot (M-1)
    \label{inequalitybalanced}
\end{equation}

\noindent because the number of elements of $(\ring^n)^{\times}$ in each $\cod\in\cod_k$ is at most $M-1\leq q^k-1$, \cite{campello2018random}. 

\subsection{Balanced set of codes from the Chinese Remainder Theorem}

We are interested in balanced sets composed of codes obtained by Construction $\pi_A$.  Let $R$ denote a commutative ring, and let $I_1, \ldots, I_s$ relatively prime ideals in $R$. According to the CRT \cite{ireland1990classical}, there exists a ring isomorphism,
\begin{equation}
\varphi: (R/\cap_{j=1}^s I_j)^n \rightarrow (R/I_1)^n\times\ldots\times(R/I_s)^n.
\label{crt_commutative}
\end{equation}

Note that $|R/\cap_{j=1}^s I_j|=|R/I_1|\cdot|R/I_2|\cdot\ldots|R/I_s|=q_1\cdot q_2\cdot\ldots\cdot q_s$. We can consider a code,
{\small \begin{equation*}
\mathcal{C} = \varphi^{-1}(\mathcal{C}_1\times \mathcal{C}_2\times \ldots\times \mathcal{C}_s), 
\end{equation*}}

\noindent where, $\mathcal{C} \subset (R/\cap_{j=1}^s I_j)^n$, and $\mathcal{C}_j \subset (R/I_j)^n, j=1, \ldots, s$.

Such isomorphism will be used in the next proposition in the case of $R=\Z, \Z[\ii]$ and $\Z[\ww]$. For the noncommutative case, $R=\Hu$, as we have seen, we also have a ring isomorphism as in (\ref{crt_commutative}), (\textit{Theorem \ref{isomorfismH}}). By abuse of notation, we denote both isomorphisms by $\varphi$.

\begin{Lema}
\label{balancedset}
Let $\mathcal{C}_{b_{k_1}}, \mathcal{C}_{b_{k_2}}, \ldots, \mathcal{C}_{b_{k_s}}$ be the balanced sets of all codes of rank $k_j$ over the finite rings $(R/I_j)^n,$ where $R=\Z,\Z[\ii], \Z[\ww]$ or $\Hu$ and $I_j, j=1, \ldots, s$ are relatively prime ideals in $R$. Then, it follows that
\begin{equation}
\mathcal{C}_{b_k} = \varphi^{-1}(\mathcal{C}_{b_{k_1}}\times \mathcal{C}_{b_{k_2}}\times \ldots \times \mathcal{C}_{b_{k_s}}),
\label{pia_balanced_set}
\end{equation}

\noindent is a balanced set of codes.
\end{Lema}

\noindent \textit{Proof:} A code $\cod$ in $\mathcal{C}_{b_k}$ has cardinality $|\mathcal{C}| = |\mathcal{C}_1| \cdot |\mathcal{C}_2| \cdot \ldots \cdot |\mathcal{C}_s|=q_1^{k_1}\cdot q_2^{k_2}\cdot \ldots\cdot q_s^{k_s}$, where $\cod_j\in\cod_{b_{k_j}}, j=1,\ldots,s$ and $k=\text{rank}(\cod)=\max\{\text{rank}(\cod_j)\}$, \cite{dougherty2017algebraic}.

Let $\ring^n = (R/\cap_{j=1}^s I_j)^n$. For any element $x$ in $(\ring^n)^{\times}$, we can find a linear bijection $T: \ring^n \rightarrow \ring^n$ such that $T(x) = (1,0,\ldots,0) = \mathbf{e}_1$. Indeed, taking $x = (x_1,\ldots,x_j,\ldots,x_n)\in(\ring^n)^{\times}$ with the $j-$th coordinate as a unit, a basis $\{x, z_1,\ldots,z_{n-1}\}$ including $x$ can be chosen, allowing us to define a linear bijection with $T(x)=\mathbf{e}_1$.

Since $T$ preserves rank and $x \in \mathcal{C}$ if and only if $\mathbf{e}_1 \in T(\mathcal{C})$ then $x$ belongs to the same number of codes in $\cod_{b_k}$ which is the number of codes $T(\cod)\in\cod_{b_k}$ that contains $\mathbf{e}_1$. Now, we can express $\mathbf{e}_1 = \varphi^{-1}(\mathbf{e}_1^{(1)},\mathbf{e}_1^{(2)},\ldots,\mathbf{e}_1^{(s)})$, where $\mathbf{e}_1^{(j)}=(1,0,\ldots,0)$ belongs to $L_j$ codes of $\mathcal{C}_{b_{k_j}}$ for $j=1,\ldots,s$. Thus, $\mathbf{e}_1$ belongs to $L = L_1 \cdot L_2 \cdot \ldots \cdot L_s$ codes of $\mathcal{C}_{b_k}$, confirming that (\ref{pia_balanced_set}) indeed defines a balanced set of codes in $\ring^n$.

\cqd

We will refer to the above set of codes $\cod_{b_k}$ as \textit{the balanced set of Construction $\pi_A$ codes} from (\ref{pia_balanced_set}).

From \textit{Lemma \ref{balancedset}} we have that this balanced set of codes satisfy the conditions of \textit{Lemma \ref{average_lema}}.

\begin{Lema}[Averaging Lemma, \cite{loeliger1997averaging}, \cite{campello2018random}] \label{average_lema}
    Let $\ring$ be a finite ring (not necessarily commutative), $g:\ring^n\rightarrow\R_+$ be a function and $\cod_{b_k}$ be a balanced set of $\ring-$linear codes of rank $k$. Then,
    \begin{equation}
        \E_{\cod_{b_k}}\left[\sum_{c\in\cod\cap (\ring^n)^{\times}} g(c)\right]\leq \dfrac{|\ring|^k-1}{|(\ring^n)^{\times}|}\sum_{x\in(\ring^n)^{\times}} g(x),
    \end{equation}
    where $\cod\in\cod_{b_k}$ and the expectation is with respect to the uniform distribution on $\cod_{b_k}$.
\end{Lema}

\subsection{Poltyrev-goodness}

Our objective is to show Construction $\pi_A$ lattices over Hurwitz integers that satisfy the requirements of the Minkowski-Hlawka Theorem. This consists of generating an ensemble of lattices using codes from a balanced set. Additionally, the reduction map used to obtain the lattice must meet a non-degeneracy condition.

We will see first, a class of Construction $\pi_A$ lattices which belong to the set of lattices obtained by reduction \cite{campello2018random}. The focus of our analysis is the isomorphism of Theorem \ref{crt2}. Similar conclusions can be derived for Construction $\pi_A$ lattices over $\Z, \Z[\ii]$ and $\Z[\ww]$. This allows to extend the next result, \textit{Proposition \ref{poltyrev-good}}, also to these cases already approached in \cite{huang2017construction} using a different technique.

The division ring of quaternions $\mathbb{H}$ can be identified with $\mathbb{R}^4$ using the natural map
\begin{equation}
\sigma(a,b,c,d) = (a + b\ii + c\jj + d\kk) \in \mathbb{H},
\label{reduction_quat}
\end{equation}

By considering this map, the Hurwitz integers corresponds to $D_4^*$, $\sigma^{-1}(\Hu) = D_4^*$, the dual of the checkerboard $\mathbb{Z}$-lattice $D_4$ in dimension 4, \cite{Con2013}. Moreover, for any rational odd prime $p$, there exists an isomorphism $\Hu/p\Hu \cong \text{M}_2(\mathbb{F}_p)$, where $\text{M}_2(\mathbb{F}_p)$ represents the ring of $2\times 2$ matrices with entries in $\mathbb{F}_p$.

Such a ring isomorphism can be defined as
{\small \begin{align*}
\rho: \Hu/p\Hu &\rightarrow \text{M}_2(\mathbb{F}_p) \nonumber\\
x_0+x_1\ii+x_2\jj+x_3\kk &\mapsto \begin{pmatrix}
x_0+x_2 a-x_3b & -x_1+x_2 b+x_3a\\
x_1+x_2 b+x_3a & x_0-x_2 a+x_3b
\end{pmatrix},
\label{construcaoAquat}
\end{align*}}

\noindent where $a,b$ are integers and $a^2+b^2+1\equiv0\pmod p$, \cite{davidoff2003elementary}.

In this way, composing these mappings enables us to obtain a reduction (surjective homomorphism) from $D_4^*$ to the ring of matrices $\text{M}_2(\mathbb{F}_p)$. By considering,
\begin{align}
\phi_p^{\Hu}: D_4^* &\stackrel{\sigma}{\longrightarrow}\Hu \stackrel{\text{mod} p}{\longrightarrow} \Hu/p\Hu \stackrel{\rho}{\longrightarrow} \text{M}_2(\mathbb{F}_p),\nonumber
\end{align}

\noindent and, given a linear code $\cod\subset M_2(\F_p)$,
\[\Lambda_p^{\Hu}(\cod)=(\phi_p^{\Hu})^{-1}(\cod),\]

\noindent represents a Hurwitz lattice obtained by reduction \cite{campello2018random}.

Consider the case where $p=\text{Nrm}(\pi)$ is an odd rational prime. We can choose $\mathcal{C}^{(1)}$ and $\mathcal{C}^{(2)}$ linear codes from $(\Hu/\Hu\pi)^n$ and $(\Hu/\Hu\overline{\pi})^n$, respectively, then $\cod\subset(\Hu/p\Hu)^n$. By abuse of notation, we will also denote by $\phi_{p}^{\Hu}$ the reduction applied component-wise to a vector in $\Hu^n\cong(D_4^*)^n$. Therefore, for Construction $\pi_A$ codes over Hurwitz integers (\textit{Definition \ref{construcao_pia_H}}), we have,
\begin{equation*}
(\phi_{p}^{\Hu})^{-1}(\cod)=\Lambda_{\pi_A}^{\Hu_p}(\mathcal{C})\subset \R^{4n}.
\end{equation*}

Let $\mathcal{C}_{b_k}$ be the balanced set of Construction $\pi_A$ codes from \textit{Lemma \ref{balancedset}}, i.e, considering $\cod^{(1)}\in\cod_{b_{k_1}}$ and $\cod^{(2)}\in\cod_{b_{k_2}}$ with $\cod_{b_{k_1}}, \cod_{b_{k_2}}$ the balanced set of all codes of rank $k_1, k_2$, respectively and $k=\max\{k_1, k_2\}$. Then, let $\la_{p}$ be the associated lattice ensembles defined as
\begin{equation}
\la_{p} = \left\{\Lambda_{p}=\beta\Lambda_{\pi_A}^{\Hu_p}(\mathcal{C}) : \mathcal{C} \in \mathcal{C}_{b_k}\right\}.
\label{random_lattices}
\end{equation}

We also remark that, since $\ker(\phi_{p_j}^{\Hu})=p\Hu$\footnote{Here, $\ker$ denotes the kernel of the linear map $\phi_p^{\Hu}$, that is, $\ker(\phi_p^{\Hu})=\{\lambda\in D_4^*; \phi_p^{\Hu}(\lambda)=0\}$.}, it follows that 
\[d_{\text{min}}(\ker(\phi_{p_j}^{\Hu}))=p>p^{1/2},\]

\noindent and the non-degeneracy condition requested in \cite{campello2018random} is valid for any prime number. Therefore, the following version from \textit{Minkowski-Hlawka Theorem}, \cite{loeliger1997averaging, campello2018random, gargava2021dense}, holds for the Construction $\pi_A$ over Hurwitz integers described above. 

\begin{Teo}
	\label{MHhurwitz}
Let $(p_j)_{j=1}^{\infty}$ be an increasing sequence of odd primes and $\cod_{b_k}$ be a balanced set of Construction $\pi_A$ codes with rank $k>n/2$. Consider the ensemble of lattices in $\R^{4n}$,
$$\la_{p_j}=\{\Lambda_{p_j}=\beta\Lambda_{\pi_A}^{\Hu_{p_j}}(\cod) : \cod \in \cod_{b_k}\},$$ 
 
\noindent where $\beta$ represents a normalization factor ensuring that all lattices in $\la_{p_j}$ have the same volume $V$. Let $f$ be a semi-admissible function\footnote{A Riemann-integrable function $f$ is considered semi-admissible if $|f(x)|\leq \dfrac{b}{(1+|x|)^{m+\delta}}, \forall x\in \mathcal{D}_f$, where $b>0$ and $\delta>0$ are positive constants.}. If we have the non-degeneracy condition,
\[d_{\text{min}}(\ker(\phi_{p_j}^{\Hu})) \geq c\cdot p_j^{1/2},\]

\noindent for some constant $c>0$. Then,
\[\lim_{p_j\rightarrow\infty} \E_{\la_{p_j}}\left[\sum_{x\in\Lambda_{p_j}(\cod)} f(\sigma^{-1}(x))\right]\leq V^{-1}\int_{\R^{4n}} f(x) dx,
\]

\noindent where, the expectation is taken over all $\Lambda_{p_j}$ within the ensemble $\la_{p_j}$.
\end{Teo}

One can now follow the proof in \cite{zamir2014lattice, loeliger1997averaging} to demonstrate that, with a high probability, the Construction $\pi_A$ generates lattices that achieve Poltyrev-goodness.

\begin{Prop}[Poltyrev-good]
\label{poltyrev-good}
	The sequence of Construction $\pi_A$ lattice in (\ref{random_lattices}) is Poltyrev-good for coding over the AWGN channel. %, that is, it is Poltyrev-good.
\end{Prop}

\section{Application to Lattice Index Coding}

Lattice index coding \cite{natarajan2015lattice} is a communication problem that exploits the presence of side information at receivers to reduce decoding complexity and improve performance over an AWGN broadcast channel. It involves encoding multiple messages into lattice codewords in such way that the receivers with some knowledge (side information) can decode more efficiently.

Consider $r$ independent messages $w_1, \ldots, w_r$ each from a finite set $W_1, \ldots, W_r$ jointly encoded into a codeword $x = f(w_1, \ldots, w_r) \in \cod$, where $\cod \subset \mathbb{R}^n$ is an $n$-dimensional constellation derived from a lattice. The received signal at receiver $l\in\{1,\ldots,L\}$ is
\[
y_l = x + z_l, \quad z_l \sim N(0, \sigma^2).
\]

Each receiver $l$ is characterised by a pair \((\mathrm{SNR}, S_l)\), where \(\mathrm{SNR}\) is the signal-to-noise ratio and \(S_l \subset \{1, \dots, r\}\) denotes the subset of messages known as side information and requires all messages \(\{w_1,\ldots, w_r\}\). Each message \(w_j\) is transmitted at a rate \(R_j = \frac{1}{n}\log_2 |W_j|\) (in bits/dim), and the total rate is \(R = \sum_{j=1}^r R_j\). The presence of side information \(w_{S_l}=\{w_j : j \in S_l\}\) at a receiver reduces the SNR required to achieve the capacity region for the AWGN network \cite{tuncel2006slepian}, since a receiver without side information requires an SNR of \(\approx2^{2R}\), while one with side information of rate \(R_{S_l} = \sum_{j \in S_l} R_j\) only requires \(\approx 2^{2(R - R_{S_l})}\), \cite{natarajan2015lattice, huang2017lattice}. 

This improvement is observed as an increase in the minimum squared Euclidean distance between valid codewords. Let \(d_0\) denote the minimum distance of the code without side information, and \(d_{S_l}\) that with side information set \(S_l\), then the squared distance gain is given by \(10 \log_{10}(d_{S_l}^2/d_0^2)\) dB. The \emph{side information gain} of a code \(\mathcal{C}\) for a index set $S_l$ is defined as \cite{natarajan2015lattice}
\[
\Gamma(\mathcal{C}, S_l) = \frac{10 \log_{10}(d_{S_l}^2/d_0^2)}{R_{S_l}} \quad \text{dB/bit/dim},
\]
and measure the SNR gain per bit/dim of side information, relative to the baseline code performance. 

Construction $\pi_A$ lattices is well suited for designing lattice index codes, particularly when the code is built from structured components over algebraic integer rings such as Hurwitz integers. 

As an example, let us construct a code over the $\Hu$ using the primes $p_1 = 3$ and $p_2 = 5$, both of which splits in $\Hu$,
\(
3 = (1+\ii+\jj)(1-\ii-\jj) = \pi_1\cdot \overline{\pi}_1 \quad \text{and} \quad 5=(1+2\ii)(1-2\ii) = \pi_2\cdot\overline{\pi}_2,
\)
\noindent and let $q=3\cdot 5 = 15$. Construction $\pi_A$ over $\Hu$ then yields the ring isomorphism,
\begin{align*}
\varphi^{-1}: \Hu/\Hu\pi_1\times\Hu/\Hu\overline{\pi}_1\times\Hu/\Hu\pi_2\times\Hu/\Hu\overline{\pi}_2\cong \Hu/15\Hu.
\end{align*}

This allows to assemble a code $\cod\subset\Hu/15\Hu$ from component codes $\cod_i\subset\Hu/\Hu\pi_i$ or $\Hu/\Hu\overline{\pi}_i$. In our example, we relabel the component codes as $\cod_1, \cod_2, \cod_3$ and $\cod_4$ and encode messages $a_i\in \cod_i, i=1,\ldots, 4$ into the codeword,
\begin{align*}
x & = \varphi^{-1}(a_1, a_2, a_3, a_4)\\
  & = (5\overline{\pi}_1a_1+5\pi_1a_2+3\overline{\pi}_2a_3+3\pi_2a_4)\mod 15\Hu.   
\end{align*}

Let, $\cod_1 = \langle \frac{1-\ii+\jj+\kk}{2},\frac{-1-\ii+\jj-\kk}{2}\rangle, \cod_2 = \langle 1+\kk \rangle, \cod_3 = \langle \frac{1+\ii+\jj+\kk}{2}\rangle$ and $\cod_4 = \langle \frac{-1+3\ii-\jj+3\kk}{2}\rangle$, linear codes over $\Hu/\Hu\pi_1,$ $ \Hu/\Hu\overline{\pi}_1,$ $ \Hu/\Hu\pi_2$ and $\Hu/\Hu\overline{\pi}_2$, respectively. Since $|\cod_1|=9$, $|\cod_2| = 3$ and $|\cod_3| = |\cod_4| = 5$, the total code size is $|\cod| = 3^3\cdot 5^2 = 675$, and as each codeword lies in a 4-dimensional real space, the rate of each message is,
\begin{align*}
R_1 = \frac{1}{2}\log_2 3, \quad  R_2 = \frac{1}{4}\log_2 3,\\
R_3 = \frac{1}{4}\log_2 5 \quad \text{and} \quad R_4 = \frac{1}{4}\log_2 5.
\end{align*}

With no side information, the receiver decodes to the nearest point in $\cod$, which has minimum Euclidean distance $d_0=1$. Now, suppose the receiver knows the first information $a_1 =0$, corresponding to side information $S=\{1\}$. The resulting subcode is,
\begin{align*}
\cod_{S} = \{5\overline{\pi}_1\cdot 0+5\pi_1a_2+3\overline{\pi}_2a_3+3\pi_2a_4 \mod15\Hu;\\
a_j\in \cod_j, j\in S^c\},
\end{align*}

\noindent which has size $|\cod_S| = 75$, side information rate of $R_S = \frac{1}{2}\log_2 3$ and minimum distance $d_S = 3$. The side information gain is given by,
\[\Gamma(\cod, S) = \dfrac{10\log_{10}(\frac{3}{1})^2}{\frac{1}{2}\log_2 3}\approx 12.0412.\]

Now considering $S = \{1, 2\}$, meaning the receiver knows $a_1 = a_2 = 0$. The subcode becomes,
\begin{align*}
\cod_{S} = \{5\overline{\pi}_1\cdot 0+5\pi_1\cdot 0+3\overline{\pi}_2a_3+3\pi_2a_4 \mod15\Hu;\\
a_j\in \cod_j, j\in S^c\},
\end{align*}

\noindent with minimum distance $d_S=3$, $|\cod_S|=25$ and the side information rate $R_S = R_1 + R_2 = \frac{1}{4}\log_2 9 + \frac{1}{4}\log_2 3 = \frac{1}{4}\log_2 27$.

In general, for any side information set $S\subset\{1,2,3,4\}$ the side information rate is given by $R_S = \frac{1}{4}\log_2\prod_{j\in S}|\cod_j|$.

The side information gain in this case is,
\[
\Gamma(\cod, S) = \dfrac{10\log_{10}(\frac{3}{1})^2}{\frac{1}{4}\log_2 27}\approx 8.0274.
\]

By repeating this same analysis for all subsets $S\subset\{1,2,3,4\}$, we obtain the complete side information gain profile in Table~\ref{side_gains}.

\begin{table}[htpb]
        \centering
        \caption{Minimum distance, message rate, and side information gain in each subcode $\cod_S$.}
        \label{side_gains}
        \begin{tabular}{c|c|c|c}
          Index set   &  Minimum distance & Message rate & $\Gamma(\cod, S)$\\
          \hline
          \hline
     $S = \{1\}$  & $3$ & $1/2 \log_2 3$ & 12.0412 \\
     $S = \{2\}$ & $3$ & $1/4 \log_2 3$ & 24.0824 \\
     $S = \{3\}$ & $\sqrt{10}$ & $1/4 \log_2 5$ & 17.2271 \\
     $S = \{4\}$ & $\sqrt{10}$ & $1/4 \log_2 5$ & 17.2271 \\
     $S = \{1,2\}$ & $3$ & $1/4 \log_2 27$ & 8.0274 \\
     $S = \{1,3\}$ & $\sqrt{30}$ & $1/4 \log_2 45$ & 10.7586 \\
     $S = \{1,4\}$ & $\sqrt{30}$ & $1/4 \log_2 45$ & 10.7586 \\
     $S = \{2,3\}$ & $\sqrt{15}$ & $1/4 \log_2 15$ & 12.0412\\
     $S = \{2,4\}$ & $\sqrt{15}$ & $1/4 \log_2 15$ & 12.0412 \\
     $S = \{3,4\}$ & $5$ & $1/4 \log_2 25$ & 12.0412 \\
     $S = \{1,2,3\}$ & $3\sqrt{5}$ & $1/4 \log_2 135$ & 9.3443 \\
     $S = \{1,2,4\}$ & $3\sqrt{5}$ & $1/4 \log_2 135$ & 9.3443 \\
     $S = \{1,3,4\}$ & $5\sqrt{3}$ & $1/4 \log_2 45$ & 9.5987 \\
     $S = \{2,3,4\}$ & $5\sqrt{3}$ & $1/4 \log_2 75$ & 12.0412 
     \vspace{.1cm}
        \end{tabular}
    \end{table}

From Table~\ref{side_gains} we observe that the side information gain $\Gamma$ depends significantly on the specific subset of messages that are known. For example, knowing message $a_2$, i.e., $S=\{2\}$, yields $\Gamma \approx 24.08$ dB/bit/dim, while knowing $a_1$, i.e., $S=\{1\}$, results in $\Gamma \approx 12.04$ dB/bit/dim. Interestingly, when both messages are known, the side information gain $\Gamma$ reduces to $\approx8.03$ dB/bit/dim. This behaviour results from the multilevel structure of Construction $\pi_A$, where each component code influences the minimum distance $d_S$, and suggest that the generator of each code must be chosen carefully. Such insights are valuable for future designs, as they suggest that prioritizing specific messages as side information can lead to more efficient decoding strategies. These properties illustrate that Construction $\pi_A$ is particularly well suited for networks where receivers have unequal access to side information.

Previous works have also explored the use of CRT for constructing lattice index codes \cite{natarajan2015lattice, huang2017lattice, huang2018lattices}. However, those approaches typically restrict the code $\cod$ to have size $q^n$, where $q$ is the product of the primes involved and $n$ is the real dimension of the space. Construction $\pi_A$ offers greater flexibility in index code design, allowing for code sizes of the form $|\cod| = |\cod_1^{(1)}|\cdot |\cod_1^{(2)}|\cdot \ldots \cdot |\cod_r^{(1)}|\cdot |\cod_r^{(2)}| = p_1^{k_1^{(1)}+k_1^{(2)}}\cdot\ldots\cdot p_r^{k_r^{(1)}+k_r^{(2)}}$, where $1\leq k_j^{(i)}\leq n$ is the rank of the linear code $\cod_j^{(i)}$, $i=1,2$ and $j=1,\ldots, r$. In the case of index coding using Construction $\pi_A$ lattices over the ring of integers a detailed investigation can be found in \cite{juliana2024indexcnmac, juliana2025uniform}.

\section{Conclusion and Perspectives}
In this work, we have extended the Construction $\pi_A$ lattice to the Hurwitz integers and analyse the potential advantages offered by such an approach.

We exploit the multilevel structure of our construction to illustrate its benefits in terms of computational complexity. It is shown in \textit{Theorem \ref{crt2}} that it is feasible to obtain an isomorphism that guarantees this extension by splitting a two-sided prime ideal into two left-prime ideals, which induces the more complete ``factorization'' given by Theorem \ref{isomorfismH} and therefore the proposed Construction $\pi_A$. The applicability of this decoding process in lattice codes with high cardinality is emphasized. In Section V we show how to construct a balanced set of codes from Construction $\pi_A$ for which the \textit{Averaging Lemma} applies and also that there exists, with high probability, a Poltyrev-good sequence of Construction $\pi_A$ lattices over Hurwitz integers.

Some promising perspectives to be considered are the extension of this construction to other maximal orders over definite quaternion algebras where it is feasible to establish an Euclidean division algorithm, \cite{fitzgerald2012norm, cerri2013euclidean} and to explore potential applications, such as compute-and-forward and other properties related to the index coding problem \cite{natarajan2015lattice, juliana2025uniform}. Additionally, it would be interesting to explore the possibility of generalizing the construction to the octonions using the maximal order of Octavian integers, which are closely related to the $E_8$ lattice.

%\appendices
%\section{Proof of the First Zonklar Equation}
%Appendix one text goes here.

% you can choose not to have a title for an appendix
% if you want by leaving the argument blank
%\section{}
%Appendix two text goes here.

% use section* for acknowledgment

%\section*{Acknowledgment}
%This work is partially supported by Brazilian foundations Coordination for the Improvement of Higher Education Personnel (CAPES -- Financial Code 001), FAPESP (2020/09838 -- 0).

% Can use something like this to put references on a page
% by themselves when using endfloat and the captionsoff option.
%\ifCLASSOPTIONcaptionsoff
%  \newpage
%\fi

% trigger a \newpage just before the given reference
% number - used to balance the columns on the last page
% adjust value as needed - may need to be readjusted if
% the document is modified later
%\IEEEtriggeratref{15}
% The "triggered" command can be changed if desired:
%\IEEEtriggercmd{\enlargethispage{-1cm}}

% references section

% can use a bibliography generated by BibTeX as a .bbl file
% BibTeX documentation can be easily obtained at:
% http://mirror.ctan.org/biblio/bibtex/contrib/doc/
% The IEEEtran BibTeX style support page is at:
% http://www.michaelshell.org/tex/ieeetran/bibtex/
\bibliographystyle{IEEEtran}
% argument is your BibTeX string definitions and bibliography database(s)
\bibliography{bibliography}

% Generated by IEEEtran.bst, version: 1.14 (2015/08/26)
\begin{thebibliography}{10}
\providecommand{\url}[1]{#1}
\csname url@samestyle\endcsname
\providecommand{\newblock}{\relax}
\providecommand{\bibinfo}[2]{#2}
\providecommand{\BIBentrySTDinterwordspacing}{\spaceskip=0pt\relax}
\providecommand{\BIBentryALTinterwordstretchfactor}{4}
\providecommand{\BIBentryALTinterwordspacing}{\spaceskip=\fontdimen2\font plus
\BIBentryALTinterwordstretchfactor\fontdimen3\font minus \fontdimen4\font\relax}
\providecommand{\BIBforeignlanguage}[2]{{%
\expandafter\ifx\csname l@#1\endcsname\relax
\typeout{** WARNING: IEEEtran.bst: No hyphenation pattern has been}%
\typeout{** loaded for the language `#1'. Using the pattern for}%
\typeout{** the default language instead.}%
\else
\language=\csname l@#1\endcsname
\fi
#2}}
\providecommand{\BIBdecl}{\relax}
\BIBdecl

\bibitem{erez2004achieving}
U.~Erez and R.~Zamir, ``Achieving 1/2 log (1+{SNR}) on the {AWGN} channel with lattice encoding and decoding,'' \emph{IEEE Transactions on Information Theory}, vol.~50, no.~10, pp. 2293--2314, 2004.

\bibitem{erez2005lattices}
U.~Erez, S.~Litsyn, and R.~Zamir, ``Lattices which are good for (almost) everything,'' \emph{Proceedings 2003 IEEE Information Theory Workshop (Cat. No.03EX674)}, pp. 271--274, 2003.

\bibitem{huang2017construction}
Y.-C. Huang and K.~R. Narayanan, ``Construction $\pi _{A}$ and $\pi _{D}$ lattices: Construction, goodness, and decoding algorithms,'' \emph{IEEE Transactions on Information Theory}, vol.~63, no.~9, pp. 5718--5733, 2017.

\bibitem{huang2018lattices}
Y.-C. Huang, K.~R. Narayanan, and P.-C. Wang, ``Lattices over algebraic integers with an application to compute-and-forward,'' \emph{IEEE Transactions on Information Theory}, vol.~64, no.~10, pp. 6863--6877, 2018.

\bibitem{huang2018layered}
Y.-C. Huang, Y.~Hong, E.~Viterbo, and L.~Natarajan, ``Layered space-time index coding,'' \emph{IEEE Transactions on Information Theory}, vol.~65, no.~1, pp. 142--158, 2019.

\bibitem{huang2017golden}
Y.-C. Huang, Y.~Hong, and B.~Viterbo, ``Golden-coded index coding,'' in \emph{2017 IEEE International Symposium on Information Theory (ISIT)}, 2017, pp. 2548--2552.

\bibitem{oggier2006perfect}
F.~Oggier, G.~Rekaya, J.-C. Belfiore, and E.~Viterbo, ``Perfect space–time block codes,'' \emph{IEEE Transactions on Information Theory}, vol.~52, no.~9, pp. 3885--3902, 2006.

\bibitem{elia2006explicit}
P.~Elia, K.~Kumar, S.~Pawar, P.~Kumar, and H.-F. Lu, ``Explicit space–time codes achieving the diversity–multiplexing gain tradeoff,'' \emph{IEEE Transactions on Information Theory}, vol.~52, no.~9, pp. 3869--3884, 2006.

\bibitem{alamouti1998}
S.~Alamouti, ``A simple transmit diversity technique for wireless communications,'' \emph{IEEE Journal on Selected Areas in Communications}, vol.~16, no.~8, pp. 1451--1458, 1998.

\bibitem{hollanti2008maximal}
C.~Hollanti, J.~Lahtonen, and H.-f. Lu, ``Maximal orders in the design of dense space-time lattice codes,'' \emph{IEEE Transactions on Information Theory}, vol.~54, no.~10, pp. 4493--4510, 2008.

\bibitem{vehkalahti2009densest}
R.~Vehkalahti, C.~Hollanti, J.~Lahtonen, and K.~Ranto, ``On the densest mimo lattices from cyclic division algebras,'' \emph{IEEE Transactions on Information Theory}, vol.~55, no.~8, pp. 3751--3780, 2009.

\bibitem{cintya2020maximal}
\BIBentryALTinterwordspacing
C.~W. de~Oliveira~Benedito, C.~Alves, N.~G. {Brasil Jr}, and S.~I.~R. Costa, ``Algebraic construction of lattices via maximal quaternion orders,'' \emph{Journal of Pure and Applied Algebra}, vol. 224, no.~5, p. 106221, 2020. [Online]. Available: \url{https://www.sciencedirect.com/science/article/pii/S0022404919302282}
\BIBentrySTDinterwordspacing

\bibitem{alves2015lattices}
\BIBentryALTinterwordspacing
C.~Alves and J.-C. Belfiore, ``Lattices from maximal orders into quaternion algebras,'' \emph{Journal of Pure and Applied Algebra}, vol. 219, no.~4, pp. 687--702, 2015. [Online]. Available: \url{https://www.sciencedirect.com/science/article/pii/S0022404914000942}
\BIBentrySTDinterwordspacing

\bibitem{hollanti2009asymetric}
C.~Hollanti and H.-F. Lu, ``Construction methods for asymmetric and multiblock space–time codes,'' \emph{IEEE Transactions on Information Theory}, vol.~55, no.~3, pp. 1086--1103, 2009.

\bibitem{Con2013}
N.~J. Conway, J. H.;~Sloane, \emph{Sphere packings, lattices and groups}.\hskip 1em plus 0.5em minus 0.4em\relax Springer Science \& Business Media, 2013, vol. 290.

\bibitem{conway2003quaternions}
J.~H. Conway and D.~A. Smith, \emph{On quaternions and octonions: their geometry, arithmetic, and symmetry}.\hskip 1em plus 0.5em minus 0.4em\relax AK Peters/CRC Press, 2003.

\bibitem{agrell2009}
\BIBentryALTinterwordspacing
E.~Agrell and M.~Karlsson, ``Power-efficient modulation formats in coherent transmission systems,'' \emph{J. Lightwave Technol.}, vol.~27, no.~22, pp. 5115--5126, Nov 2009. [Online]. Available: \url{https://opg.optica.org/jlt/abstract.cfm?URI=jlt-27-22-5115}
\BIBentrySTDinterwordspacing

\bibitem{karlsson2011}
M.~Karlsson and E.~Agrell, \emph{Power-Efficient Modulation Schemes}.\hskip 1em plus 0.5em minus 0.4em\relax New York, NY: Springer New York, 2011, pp. 219--252.

\bibitem{zetterberg1977}
L.~Zetterberg and H.~Brandstrom, ``Codes for combined phase and amplitude modulated signals in a four-dimensional space,'' \emph{IEEE Transactions on Communications}, vol.~25, no.~9, pp. 943--950, 1977.

\bibitem{welti1974}
G.~Welti and J.~Lee, ``Digital transmission with coherent four-dimensional modulation,'' \emph{IEEE Transactions on Information Theory}, vol.~20, no.~4, pp. 497--502, 1974.

\bibitem{natarajan2015lattice}
L.~Natarajan, Y.~Hong, and E.~Viterbo, ``Lattice index coding,'' \emph{IEEE Transactions on Information Theory}, vol.~61, no.~12, pp. 6505--6525, 2015.

\bibitem{freudenberger2017generalized}
J.~Freudenberger, D.~Rohweder, and S.~Shavgulidze, ``Generalized multistream spatial modulation with signal constellations based on {H}urwitz integers and low-complexity detection,'' \emph{IEEE Wireless Communications Letters}, vol.~7, no.~3, pp. 412--415, 2018.

\bibitem{stern2018quaternion}
S.~Stern and R.~F. Fischer, ``Quaternion-valued multi-user {MIMO} transmission via dual-polarized antennas and {QLLL} reduction,'' in \emph{25th International Conference on Telecommunications (ICT)}, 2018, pp. 63--69.

\bibitem{ugrelidze2020new}
N.~Ugrelidze, S.~Shavgulidze, and M.~Sordia, ``New four-dimensional signal constellations construction,'' \emph{IET Communications}, vol.~14, no.~10, pp. 1554--1559, 2020.

\bibitem{stern2022algorithms}
S.~Stern, C.~Ling, and R.~F. Fischer, ``Algorithms and bounds for complex and quaternionic lattices with application to mimo transmission,'' \emph{IEEE Transactions on Information Theory}, vol.~68, no.~7, pp. 4491--4517, 2022.

\bibitem{vance2011improved}
S.~Vance, ``Improved sphere packing lower bounds from {H}urwitz lattices,'' \emph{Advances in Mathematics - Advan Math}, vol. 227, 05 2011.

\bibitem{loeliger1997averaging}
H.-A. Loeliger, ``Averaging bounds for lattices and linear codes,'' \emph{IEEE Transactions on Information Theory}, vol.~43, no.~6, pp. 1767--1773, 1997.

\bibitem{forney2000sphere}
G.~Forney, M.~Trott, and S.-Y. Chung, ``Sphere-bound-achieving coset codes and multilevel coset codes,'' \emph{IEEE Transactions on Information Theory}, vol.~46, no.~3, pp. 820--850, 2000.

\bibitem{Sue2018}
S.~I.~R. Costa, F.~Oggier, A.~Campello, J.-C. Belfiore, and E.~Viterbo, \emph{Lattices applied to coding for reliable and secure communications}.\hskip 1em plus 0.5em minus 0.4em\relax Springer, 2017.

\bibitem{kositwattanarerk2015construction}
W.~Kositwattanarerk, S.~S. Ong, and F.~Oggier, ``Construction {A} of lattices over number fields and block fading (wiretap) coding,'' \emph{IEEE Transactions on Information Theory}, vol.~61, no.~5, pp. 2273--2282, 2015.

\bibitem{vehkalahti2014constructions}
R.~Vehkalahti, W.~Kositwattanarerk, and F.~Oggier, ``Constructions {A} of lattices from number fields and division algebras,'' in \emph{2014 IEEE International Symposium on Information Theory}, 2014, pp. 2326--2330.

\bibitem{hlawka1943geometrie}
E.~Hlawka, ``Zur geometrie der zahlen,'' \emph{Mathematische Zeitschrift}, vol.~49, no.~1, pp. 285--312, 1943.

\bibitem{Cas1997}
J.~W.~S. Cassels, \emph{An introduction to the geometry of numbers}.\hskip 1em plus 0.5em minus 0.4em\relax Springer Science \& Business Media, 1997.

\bibitem{zamir2014lattice}
R.~Zamir, \emph{Lattice Coding for Signals and Networks: A Structured Coding Approach to Quantization, Modulation, and Multiuser Information Theory}.\hskip 1em plus 0.5em minus 0.4em\relax Cambridge University Press, 2014.

\bibitem{poltyrev1994coding}
G.~Poltyrev, ``On coding without restrictions for the {AWGN} channel,'' \emph{IEEE Transactions on Information Theory}, vol.~40, no.~2, pp. 409--417, 1994.

\bibitem{hamilton1848}
W.~R. Hamilton, ``On quaternions; or on a new system of imaginaries in algebra,'' \emph{The London, Edinburgh and Dublin Philosophical Magazine and Journal of Science}, vol.~33, no. 219, pp. 58--60, 1848.

\bibitem{reyes2010one}
M.~L. Reyes, ``One-sided prime ideal principle for noncommutative rings,'' \emph{Journal of Algebra and Its Applications}, vol.~09, no.~06, pp. 877--919, 2010.

\bibitem{reiner1975maximal}
I.~Reiner, \emph{{Maximal Orders}}.\hskip 1em plus 0.5em minus 0.4em\relax Oxford University Press, 01 2003.

\bibitem{voight2021quaternion}
J.~Voight, \emph{Quaternion algebras}.\hskip 1em plus 0.5em minus 0.4em\relax Springer Nature, 2021.

\bibitem{ireland1990classical}
K.~Ireland and M.~Rosen, \emph{A classical introduction to modern number theory}.\hskip 1em plus 0.5em minus 0.4em\relax Springer, 1990, vol.~16.

\bibitem{natarajan2015lattice2}
L.~Natarajan, Y.~Hong, and E.~Viterbo, ``Lattice index coding for the broadcast channel,'' in \emph{2015 IEEE Information Theory Workshop (ITW)}.\hskip 1em plus 0.5em minus 0.4em\relax IEEE, 2015, pp. 1--5.

\bibitem{huang2017lattice}
Y.-C. Huang, ``Lattice index codes from algebraic number fields,'' \emph{IEEE Transactions on Information Theory}, vol.~63, no.~4, pp. 2098--2112, 2017.

\bibitem{matsumine2018construction}
T.~Matsumine, B.~M. Kurkoski, and H.~Ochiai, ``Construction {D} lattice decoding and its application to {BCH} code lattices,'' in \emph{2018 IEEE Global Communications Conference}, 2018, pp. 1--6.

\bibitem{davey1998}
M.~Davey and D.~MacKay, ``Low density parity check codes over gf(q),'' in \emph{1998 Information Theory Workshop (Cat. No.98EX131)}, 1998, pp. 70--71.

\bibitem{zhou2016}
X.~Zhou and C.~Xie, \emph{Multidimensional Optimized Optical Modulation Formats}.\hskip 1em plus 0.5em minus 0.4em\relax Wiley, 2016, pp. 13--64.

\bibitem{stern2019hurwitz}
S.~Stern, F.~Frey, J.~K. Fischer, and R.~F.~H. Fischer, ``Two-stage dimension-wise coded modulation for four-dimensional hurwitz-integer constellations,'' in \emph{SCC 2019; 12th International ITG Conference on Systems, Communications and Coding}, 2019, pp. 1--6.

\bibitem{frey2020}
F.~Frey, S.~Stern, J.~K. Fischer, and R.~F.~H. Fischer, ``Two-stage coded modulation for hurwitz constellations in fiber-optical communications,'' \emph{Journal of Lightwave Technology}, vol.~38, no.~12, pp. 3135--3146, 2020.

\bibitem{agrell2019database}
\BIBentryALTinterwordspacing
E.~Agrell. (2019) Database of sphere packings. Last accessed 10 May 2025. [Online]. Available: \url{https://codes.se/packings/}
\BIBentrySTDinterwordspacing

\bibitem{forney1989constellation}
G.~Forney and L.-F. Wei, ``Multidimensional constellations. i. introduction, figures of merit, and generalized cross constellations,'' \emph{IEEE Journal on Selected Areas in Communications}, vol.~7, no.~6, pp. 877--892, 1989.

\bibitem{nebe2014database}
\BIBentryALTinterwordspacing
G.~Nebe and N.~Sloane. (2014) A catalogue of lattices. Last accessed 10 May 2025. [Online]. Available: \url{https://www.math.rwth-aachen.de/~Gabriele.Nebe/LATTICES/index.html#4D}
\BIBentrySTDinterwordspacing

\bibitem{dougherty2017algebraic}
S.~T. Dougherty, \emph{Algebraic coding theory over finite commutative rings}.\hskip 1em plus 0.5em minus 0.4em\relax Springer, 2017.

\bibitem{hartley1970modules}
B.~Hartley and T.~O. Hawkes, \emph{Rings, Modules and Linear Algebra}.\hskip 1em plus 0.5em minus 0.4em\relax Chapman and Hall, London, 1970.

\bibitem{campello2018random}
A.~Campello, ``Random ensembles of lattices from generalized reductions,'' \emph{IEEE Transactions on Information Theory}, vol.~64, no.~7, pp. 5231--5239, 2018.

\bibitem{davidoff2003elementary}
G.~P. Davidoff, P.~Sarnak, and A.~Valette, \emph{Elementary number theory, group theory, and Ramanujan graphs}.\hskip 1em plus 0.5em minus 0.4em\relax Cambridge university press Cambridge, 2003, vol.~55.

\bibitem{gargava2021dense}
N.~Gargava and V.~Serban, ``Dense packings via lifts of codes to division rings,'' \emph{IEEE Transactions on Information Theory}, vol.~69, no.~5, pp. 2860--2873, 2023.

\bibitem{tuncel2006slepian}
E.~Tuncel, ``Slepian-wolf coding over broadcast channels,'' \emph{IEEE Transactions on Information Theory}, vol.~52, no.~4, pp. 1469--1482, 2006.

\bibitem{juliana2024indexcnmac}
J.~G.~F. Souza and S.~I.~R. Costa, ``Lattice index coding from construction $\pi_{A}$ lattices,'' \emph{Proceeding Series of the Brazilian Society of Computational and Applied Mathematics}, vol.~11, no.~1, 2025.

\bibitem{juliana2025uniform}
\BIBentryALTinterwordspacing
------, ``Achieving uniform side information gain with multilevel lattice codes over the ring of integers,'' 2025, to appear on IEEE Communications Letter. [Online]. Available: \url{https://arxiv.org/abs/2501.14921}
\BIBentrySTDinterwordspacing

\bibitem{fitzgerald2012norm}
R.~W. Fitzgerald, ``Norm euclidean quaternionic orders,'' \emph{Integers Journal}, vol.~12, no.~2, pp. 197--208, 2012.

\bibitem{cerri2013euclidean}
J.-P. Cerri, J.~Chaubert, and P.~Lezowski, ``Euclidean totally definite quaternion fields over the rational field and over quadratic number fields,'' \emph{International Journal of Number Theory}, vol.~09, no.~03, pp. 653--673, 2013.

\end{thebibliography}

% biography section
 
% If you have an EPS/PDF photo (graphicx package needed) extra braces are
% needed around the contents of the optional argument to biography to prevent
% the LaTeX parser from getting confused when it sees the complicated
% \includegraphics command within an optional argument. (You could create
% your own custom macro containing the \includegraphics command to make things
% simpler here.)
%\begin{IEEEbiography}[{\includegraphics[width=1in,height=1.25in,clip,keepaspectratio]{mshell}}]{Michael Shell}
% or if you just want to reserve a space for a photo:

%\begin{IEEEbiography}{Michael Shell}
%Biography text here.
%\end{IEEEbiography}

% if you will not have a photo at all:
%\newpage

\begin{IEEEbiographynophoto}{Juliana G. F. Souza}
(Student Member, IEEE)  was born in Belém, Pará, Brazil. She received the Mathematics degree from the University of Pará (UFPA), in 2010, the M.Sc. degree in Applied and Computational Mathematics from the University of Campinas (Unicamp), São Paulo, Brazil, 2019 and she is currently a Ph.D. student in Applied Mathematics at Unicamp. She held a short-term internship at the Electrical and Electronic Engineering Department, Imperial College London, in 2023. Her research interests include information theory and lattice coding.
\end{IEEEbiographynophoto}

%\vfill

% insert where needed to balance the two columns on the last page with
% biographies

%\newpage

\begin{IEEEbiographynophoto}{Sueli I. R. Costa}
(Member, IEEE) a Professor at the Institute of Mathematics, Statistics and Computer Science, University of Campinas, Brazil, received her Ph.D. from the same university and had her postdoctoral studies at the Institute for Advanced Study, Princeton. Her professional activities have included teaching, supervising graduate students and post-doctoral researchers, coordinating a graduate program, several short term visits to other universities and centers and coordinating research development projects. She has served as a co- chair of the 2011 IEEE ITW, of the 2018 Latin American Week on Coding and Information, as the IEEE Information Society Brazil Chapter chair (2015-2021), as a member of the IEEE Hamming Medal Committee (2019-2021) and as a member of the IEEE Information Theory Society Awards Committee (2023-2024). Her research topics of interest include lattice codes in communications, discrete and continuous spherical codes, coding for storage and information geometry.
\end{IEEEbiographynophoto}

\begin{IEEEbiographynophoto}{Cong Ling}
(Member, IEEE) received the bachelor’s and master’s degrees from Nanjing Institute of Communications Engineering, China, in 1995 and 1997, respectively, and the Ph.D. degree from Nanyang Technological University, Singapore, in 2005. He is currently a Reader (equivalent to Professor/Associate Professor) with the Electrical and Electronic Engineering Department, Imperial College London. He is also a member of the Academic Centre of Excellence in Cyber Security Research, Imperial College, and an affiliated member of the Institute of Security Science and Technology, Imperial College. Before joining Imperial College, he had been with the Faculties at Nanjing Institute of Communications Engineering and King College. He visited The Hong Kong University of Science and Technology as a Hong Kong Telecom Institute of Information Technology (HKTIIT) fellow in 2009. He has been an Associate Editor in multiterminal communications and lattice coding for IEEE TRANSACTIONS ON COMMUNICATIONS and an Associate Editor of IEEE TRANSACTIONS ON VEHICULAR TECHNOLOGY and on the program committees of several international conferences, including IEEE Information Theory Workshop, Globecom, and ICC. 
\end{IEEEbiographynophoto}

% You can push biographies down or up by placing
% a \vfill before or after them. The appropriate
% use of \vfill depends on what kind of text is
% on the last page and whether or not the columns
% are being equalized.

\vfill

% Can be used to pull up biographies so that the bottom of the last one
% is flush with the other column.
%\enlargethispage{-10cm}

% that's all folks
\end{document}